# The Golden Age of Statistical Graphics

**Michael Friendly**

*Abstract.* Statistical graphics and data visualization have long histories, but their modern forms began only in the early 1800s. Between roughly 1850 and 1900 (±10), an explosive growth occurred in both the general use of graphic methods and the range of topics to which they were applied. Innovations were prodigious and some of the most exquisite graphics ever produced appeared, resulting in what may be called the "Golden Age of Statistical Graphics."

In this article I trace the origins of this period in terms of the infrastructure required to produce this explosive growth: recognition of the importance of systematic data collection by the state; the rise of statistical theory and statistical thinking; enabling developments of technology; and inventions of novel methods to portray statistical data. To illustrate, I describe some specific contributions that give rise to the appellation "Golden Age."

## 1. INTRODUCTION

Data and information visualization is concerned with showing quantitative and qualitative information, so that a viewer can see patterns, trends or anomalies, constancy or variation, in ways that other forms—text and tables—do not allow. Today, statistical graphs and maps are commonplace, and include time-series graphs of economic indicators, pie- and bar-charts of government spending, maps of election results (e.g., the "red" and "blue" maps of U.S. Presidential races), maps of disease incidence or outbreak (perhaps related visually to potential causes) and so forth.

"New" graphical methods are frequently proposed to help convey to the eyes an increasingly complex range, size and scope of the data of modern science and statistical analysis. However, this is often done without an appreciation or even understanding of their antecedents. As I hope will become apparent here, many of our "modern" methods of statistical graphics have their roots in the past and came to fruition in a particular period of time.

The graphic display of data has a very long history (Friendly and Denis, 2000; Friendly, 2008), but the age of modern statistical graphs and maps only began around the beginning of the 19th century. In statistical graphics, William Playfair [1759–1823] invented the line graph and bar chart (Playfair, 1786), followed by the pie chart and circle graph (Playfair, 1801). Statistical maps have their modern origin in the use of isolines, showing curves of constant value (wind directions and magnetic declination), by Edmund Halley [1656–1742] (Halley, 1701). Another map technique, the use of continuous shading (from light to dark) to show the geographic distribution of regional values (literacy in France), was first used by Baron Charles Dupin [1784–1873] (Dupin, 1826).

With these innovations in design and technique, the first half of the 19th century was an age of enthusiasm for graphical display (Funkhouser, 1937;

*Michael Friendly is Professor, Psychology Department, York University, Toronto, ON, M3J 1P3 Canada e-mail: friendly@yorku.ca.*

Palsky, 1996) and witnessed explosive growth in statistical graphics and thematic mapping, at a rate which would not be equaled until recent times. The period from about 1840–1850 until 1900–1910 saw this rapid growth continue, but did something more.

In the latter half of the 19th century, youthful enthusiasm matured, and a variety of developments in statistics, data collection and technology combined to produce a "perfect storm" for data graphics. The result was a qualitatively distinct period which produced works of unparalleled beauty and scope, the likes of which would be hard to duplicate today. I argue that this period deserves to be recognized and named; I call it the "Golden Age of Statistical Graphics." By the end of this period, statistical graphics had become mainstream; yet, paradoxically, the enthusiasm, innovation and beauty of the Golden Age would soon die out.

To give some initial (and graphic) sense of these periods in the history of data visualization and the growth and decline in the innovations referred to above, I consider Figure 1, showing the time distribution of 260 items from the database of the Milestones Project (Friendly, 2005). This is a comprehensive catalog of statistical and graphical developments which are considered to be significant events ("milestones") in this history. The dashed lines and labels for periods reflect a convenient parsing of this history described elsewhere in detail (Friendly, 2008). The fringe marks at the bottom (a rug plot) show the discrete milestone events; the smoothed curve shows a nonparametric kernel density estimate of the relative frequency of these events over time. The rapid rise in this curve in the 1800s, followed by a steep decline in the early 1900s, gives graphic form to the developments I seek to explain here.

This article traces the origins of the Golden Age of Statistical Graphics in terms of the infrastructure required to produce this rapid growth: recognition of the importance of systematic data collection by the state; the rise of statistical theory and statistical thinking; enabling developments of technology; and inventions of novel methods to portray statistical data. Some individual contributions to the Golden Age and the role of government-sponsored statistical albums in fostering graphical excellence are then illustrated. The scope of this paper is largely restricted to developments leading up to, and those that follow directly from, the Golden Age. The final historical section discusses the reasons for the end of the Golden Age, leading to a period that can be called the Modern Dark Ages of statistical graphics, and also some later contributions to renewed interest and current practice that relate to this period. More general accounts of this history are given elsewhere (Friendly, 2005, 2008).

### 1.1 What is an Age? What Makes One Golden?

Before proceeding, it is helpful to clarify the idea of a Golden Age to demonstrate why the developments I have just summarized, and those I will detail below, deserve to be recognized as the Golden Age of Statistical Graphics.

We think of an *Age* as a period in history not necessarily with sharp boundaries, but appearing qualitatively and perhaps quantitatively distinct from the periods before and after. The term *Golden Age* originated from early Greek and Roman poets who used it to refer to a time when mankind was pure and lived in a utopia; but more generally it is used to refer to some recognizable period in a field or region where great tasks were accomplished.

Some examples of Golden Ages in this sense are: (a) the Golden Age of Athens under Pericles between the end of the Persian War (448 BCE) and the beginning of the Peloponnesian Wars (404 BCE), a relative high-point in the development of politics and civil society, architecture, sculpture and theater; (b) the Golden Age of Islam (750–1258) from the solidification of the Islamic caliphate to the sack of Baghdad by the Moguls, during which there were great advances in the arts, science, medicine and mathematics; (c) the Golden Age of England, 1558–1603 under Elizabeth I, a peak in Renaissance literature, poetry and theater. Such periods often end with some turning-point event(s).

Statistically, one can think of a Golden Age as a local maximum in some distribution over history. From Figure 1 we can see that the number of events and innovations recorded as milestones in the history of data visualization grew rapidly throughout the 1800s but then suffered a down-turn toward the end of the century. This is just one quantitative indicator of the development of graphical methods in the Golden Age.

Historians of statistical graphics and thematic cartography, including Funkhouser (1937), Robinson (1982) and Palsky (1996), have all referred to this period, plus or minus one or two decades, as one that deserves a special designation. Funkhouser (1937), Chapter 5, page 329, and Palsky (1996), Chapter 4, refer to the last half of the 19th century as an "Age of Enthusiasm" for graphics. It certainly was that, but, as I hope to show below, the grander title of "Golden Age" is certainly appropriate. It is worth quoting Funkhouser (1937), page 330, at length here:

> The period from 1860 to 1890 may be called the golden age of graphics, for it was marked by the unrestrained enthusiasm not only of statisticians but of government and municipal authorities, by the eagerness with which the possibilities and problems of graphic representation were debated and by the graphic displays which became an important adjunct of almost every kind of scientific gathering. During this period the method was officially recognized by government agencies and became a feature of official publications. Here also is found the first reference to the graphic method as a universal language together with the opinion of more sober statisticians that the method was running away with itself.

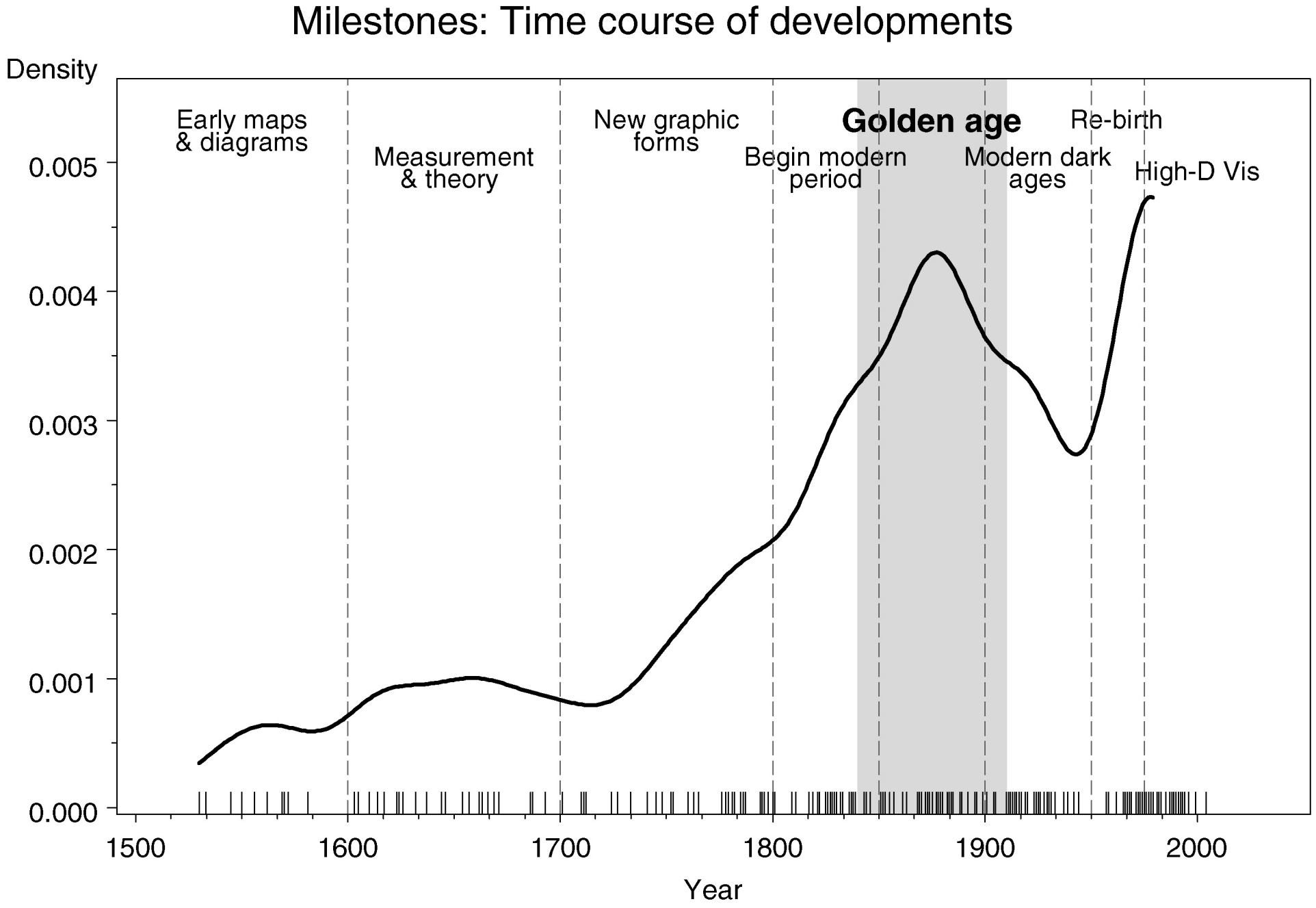

FIG. 1. *The time distribution of events considered milestones in the history of data visualization, shown by a rug plot and density estimate. The density estimate is based on* $n = 260$ *significant events in the history of data visualization from 1500–present, and uses the Sheather–Jones (1991) plug-in estimator for bandwidth selection. In the plot, the density curve is truncated at 1985 to avoid end effects. The developments in the highlighted period, from roughly 1840–1910, comprise the subject this paper seeks to explain.*

## 2. PRELUDES TO THE GOLDEN AGE

It is not sufficient to simply describe and illustrate the intellectual and graphical accomplishments of this period I call the Golden Age of Statistical Graphics, as I do in Section 3. Rather, to fully appreciate how this period arose, it is necessary to understand the infrastructure required to produce this explosive growth—in quantity, range of application, and beauty—in the last half of the 19th century. As well, in writing this account I have been struck by how the interplay among these elements—data, statistical theory, technological advances, and inventions in techniques of visual representation—contributed in *combination* to the advances of this period.

The topics described below are developed only insofar as to illuminate their contributions to the development of graphic representations of data. In particular, important contributions of many individuals in these areas are omitted in the interest of brevity.

### 2.1 Statistics: Numbers of the State

The word *statistics* derives from the post-medieval latin *statisticum* ("of the state") and was introduced around 1750 as "statistik"[1] by Gottfried Achenwall [1719–1772] to refer to the collection and tabulation of numbers about the state (Achenwall, 1749). The coining of this word recognized the birth of interest in data about human populations that had begun nearly a century before.

In the 1660s, John Graunt's *Observations upon Bills of Mortality* (Graunt, 1662) demonstrated how data on births and deaths could be used to inform the state about matters related to population dynamics and human capital: age-related mortality, the ability to raise an army or to extract taxes. These ideas were quickly framed as "political arithmetic" by Sir William Petty (1665, 1690) and attracted widespread attention in the kingdoms of Europe.

By the mid-1700s, the importance of measuring and analyzing population distributions and the idea that state policies could encourage wealth through population growth were established, most notably by the Prussian Johan Peter Süssmilch (1741), who advocated expansion of governmental collection of population statistics.

At the start of the 19th century there occurred a burst of interest in numerical data on a much wider variety of topics, as detailed below. In response, the first statistical societies were formed, beginning with the Statistical Society of London, organized by Adolphe Quetelet and Charles Babbage in February 1834. Initially, the Society set its objective as "the collection and classification of all facts illustrative of the present condition and prospects of society, especially as it exists in the British Dominions." In this statement, they implicitly eschewed the analysis and interpretation of such data, fearing to become involved in political discussions regarding crime, social and medical conditions. As a result, these early statisticians called themselves "statists," and confined themselves largely to the tabular presentation of statistical "facts." This tabular orientation had already begun to change elsewhere in Europe, but graphics would not be officially welcomed into the Royal Statistical Society until the Jubilee meeting of 1885 (Marshall, 1885).

[1]The first known uses of the latin form occur in the 1670s; for example, in the 1672 work under the pseudonym Helenus Politanus, *Microscopium statisticum: quo status imperii Romano-Germanici* .... In France, the introduction of the word "statistique" is often ascribed to a 1665 memoire by Claude Bouchu, administrator of Bourgogne, titled *Déclaration des biens, charges, dettes et statistique des communautés de la généralité de Dijon* .... It now seems more likely that the word "statistique" was later added to the title by an archivist (Pepin, 2005), but the word did appear in French dictionaries around 1700.

## 2.2 An Avalanche of Social Numbers

*Criminal statistics becomes as empirical and accurate as the other observational sciences when one restricts oneself to the best-observed facts and groups them in such a way as to minimize accidental variation. General patterns then appear with such regularity that it is impossible to attribute them to random chance. ... We are forced to recognize that the facts of the moral order are subject, like those of the physical order, to invariable laws.* —Guerry (1833), page 10ff.

Collection of data on population and economic conditions (imports, exports, etc.) became widespread in European countries by the beginning of the 19th century. However, there were few data relating to social issues. This was to change dramatically in the period from about 1820 to 1830, and would impel the application of graphical methods to important social issues.

By 1822, some initial data on the levels of school instruction by region or political division for some countries in Europe were published by Adriano Balbi in *Essai statistique sur le royaume de Portugal et d'Algarve, comparé aux autres Etats de l'Europe* (Balbi, 1822). In Paris, J. B. Joseph Fourier and Frédéric Villot began a massive tabulation of births, marriages, deaths (by cause), admission to insane asylums (by age, sex, affliction), causes of suicides, and so forth, published as the *Recherches statistiques sur la ville de Paris et le départementde la Seine* starting in 1821 (Fourier et al., 1860); see also Hacking (1990), pages 73–77.

Following Napoleon's defeat there was widespread upheaval in France, resulting in explosive growth in Paris, large-scale unemployment and a popular perception, particularly among the upper classes, that crime was on the rise. Then (as now), newspapers carried lurid accounts of robberies, murders and other crimes committed by the "dangerous classes" (Chevalier, 1958). Perhaps in response, the French Ministry of Justice initiated the *Compte général*—the first comprehensive, centralized, *national* system of crime reporting, collected quarterly from every department and recording the details of *every* criminal charge laid before the French courts: age, sex, and occupation of the accused, the nature of the charge, and the outcome in court.

Very quickly, these data were portrayed in graphic forms (maps and diagrams) through which they were used to shed some light on social issues, in ways that the tables of the British statists could not. Baron Charles Dupin used data similar to Balbi's (Dupin, 1826, 1827), to draw a map where each department was shaded according to the level of educational instruction (represented by the number of male school children per unit population), using tints of varying darkness to depict degrees of ignorance. This was the first modern statistical map (now called a choropleth map), and showed a dramatic result: a clear demarcation could be seen between the north and south of France along a line running southwest from about Saint-Malo in Brittany to near Geneva.[2]

At this time, André-Michel Guerry [1802–1866] was a young lawyer with a penchant for numbers. In 1827 he began to work with the data from the *Compte général* in the course of his duties with the Ministry of Justice. He became so fascinated by these data that he abandoned active practice in law to devote himself to their analysis, a task he would pursue until his death in 1866; see Friendly (2007) for his full story.

Here, it suffices to illustrate this development with the first truly high-impact graphic based on such data. Guerry and Balbi (1829) wanted to show the *relation* of crimes against persons and crimes against property with levels of instruction. Was it true that crime was low where education was high? Did personal crime and property crime show the same or different relations? To answer this, they composed a single graphic (see Friendly, 2007, Figure 2[3]) containing three maps representing these data by Dupin's method, so crimes and instruction could readily be compared. The result was rather striking: neither personal crime nor property crime seemed to have any direct relation to level of instruction, and these two types of crimes seemed inversely related across the departments of France.

[2]The "Saint-Malo–Geneva line," a sharp cleavage between *France du Nord et Midi*, was later reified as a contrast of *France éclairée* vs. *France obscure*. This difference would generate much debate about causes and circumstances through the end of the 19th century.

[3]Available at www.math.yorku.ca/SCS/Gallery/images/guerry/guerry-balbi-600s.jpg.

This combination of even rudimentary graphical methods with the systematic collection of data concerning important issues would propel the development and use of graphics into the Golden Age, even absent statistical theory to simplify results and to test relations shown graphically. Happily, statistical ideas and methods were simultaneously on the rise, a topic I turn to next.

## 2.3 Statistical Theory and Statistical Thinking

*I know of scarcely anything so apt to impress the imagination as the wonderful form of cosmic order expressed by the "Law of Frequency of Error." The law would have been personified by the Greeks and deified, if they had known of it. ... It is the supreme law of Unreason.* —Sir Francis Galton, *Natural Inheritance*, London: Macmillan, 1889.

Modern statistical theory had its origins in a diverse collection of practical and theoretical problems. These included:

- games of chance, which gave rise to initial statements of a theory of probability in the 1650s by Blaise Pascal and Pierre de Fermat and quickly appeared in a textbook by Christiaan Huygens (Huygens, 1657);
- astronomical and geodetic observations, used to calculate the orbits of planets and comets and determine the shape of the earth, with the practical goal of enabling accurate navigation at sea.

This last set of problems, from the early 1700s to the early 1800s, would occupy the best mathematical minds of the century and would give rise to the fundamental ideas from which statistics grew as a discipline. First, the idea of combining observations to give a "best" estimate of some true value in the face of errors of measurement (Roger Cotes, 1722; Tobias Mayer, 1750) led to the development of the method of least squares between 1780 and 1810 (contributions by Pierre Laplace, Adrien Legendre, Carl Friedrich Gauss). A second fundamental idea, that stemmed as much from probability theory, was that of assessing the accuracy or uncertainty of estimated values (Jacob Bernoulli, Abraham De Moivre). This led to the development of the binomial distribution (De Moivre, 1718), the central limit theorem (Laplace, 1812) and the normal distribution (De Moivre, 1738).

These theoretical ideas had little impact on the development of statistical graphics, at least initially.

What *did* have a substantial consequence in this regard was the rise of statistical thinking (Porter, 1986) and extension of these ideas to the human and social realms which occurred after 1820 with the explosion of social data described above (Section 2.2).

Among the most important extensions was the transformation of the "Law of Frequency of Error" into the "Supreme Law of Unreason," so extravagantly extolled by Galton in the quotation above, into a scientific and philosophical basis for the study of human characteristics. In the calculus of astronomical observations, the *central value* (or least squares estimate) was of primary concern. The reinterpretation pertaining to the distribution of human characteristics, particularly in relation to the normal distribution, made the study of laws of *variability* equally important.

This transformation is best embodied in the works of Adolphe Quetelet [1796–1874], perhaps the first to use the normal distribution other than as a law of error, whose concept of the "average man," in physical and moral characteristics, stimulated statistical and social thought throughout Europe, and André-Michel Guerry who showed that these moral characteristics (crime, suicide, literacy and so forth) exhibited *stability* in their central value over time, yet systematic *variability* over place and circumstances, and so could be taken to imply the existence of social laws, akin to the laws of physical science.

The last significant developments of statistical thinking in the 19th century were the concepts of bivariate relations, linear regression and the bivariate normal distribution developed first in a general way by Francis Galton (1886) in his studies of heredity and then used by Karl Pearson (1896) in a rigorous treatment of correlation and regression. These concepts came too late to have much impact on the statistical graphics of the Golden Age, even though questions of *relations* between variables were important ones in the substantive areas outlined in Section 2.2.

For example, the essential idea of the scatterplot was first described by John F. W. Herschel [1792–1871] (Herschel, 1833). Remarkably, he also described the idea of smoothing a bivariate relation to make the trend more apparent and allow prediction and interpolation (Friendly and Denis, 2005; Hankins, 2006). Yet, Guerry (1833, 1864) and the other moral statisticians who followed him (e.g., Fletcher, 1847) never thought to make scatterplots of, say, crime or suicide against education to visualize the relation *directly*, despite the fact that this was their main goal. Statistical theory and statistical thinking enabled the contributions of the Golden Age, but were often necessary antecedents.

### 2.4 Technology

Advances in technology served a purely practical, but important role in the development of graphics during the 19th century, by allowing images to be produced more quickly and cheaply and disseminated more widely, and also by allowing statistical data to be collected more easily and summarized.

2.4.1 *Lithography and color printing.* *At the time, the effect of lithography in the field of duplication was as great as has been the introduction in our time of rapid-copying techniques such as Xerox* — (Robinson, 1982, page 57).

The difficulties of printing and reproduction have always been impediments to the use and dissemination of statistical graphics; indeed, even today technical capabilities and economics of publication place limitations on how data graphics can be shown. In the period leading up to the Golden Age, thematic maps and diagrams were printed by means of copperplate engraving, whereby an image could be incised on a soft copper sheet, then inked and printed. In the hands of master engravers and printers, copperplate technology could easily accommodate fine lines, small lettering, stippled textures and so forth. The resulting images were far superior to those produced by previous woodcut methods, but copperplate was slower, more costly, and required different print runs. The graphs in Playfair's major works (Playfair, 1786, 1801), for example, were printed via copperplate and hand-colored, and therefore printed in limited numbers.

Lithography, a chemical process for printing invented in 1798 by Aloys Senefelder [1771–1843], allowed for much longer print runs of maps and diagrams than engraving, was far less expensive, and also made it easier to achieve fine tonal gradation in filled areas. The method began to be widely used after the publication of English (Senefelder, 1819), French and German editions of Senefelder's *Complete Course of Lithography.* The quotation above from Robinson 1982 attests to the influence of lithography on the dissemination of statistical maps and diagrams.

By around 1850, lithographic techniques were adapted to color printing, making the use of color much more frequent, but, more importantly, permitting color to be used as an important perceptual feature in the design of thematic maps and statistical diagrams. It is no accident that the widespread use of color is a major characteristic of the Golden Age.

2.4.2 *Automatic recording and calculation.* Other technological developments also contributed to the birth of the Golden Age. Graphic recording devices—instruments that turn a time-varying phenomenon into a graphic record—have a history (Hoff and Geddes, 1962) that dates back to antiquity. The modern history of this technology probably starts with a publication by Alexander Keith (1800) in the newly formed *Journal of Natural Philosophy*, describing the idea for continuous pen-recorders of temperature and barometric pressure, creating automatic weather graphs, but also tracking the maximum and minimum over time. By 1822, James Watt (1822) (with John Southern) published a description of the "Watt Indicator," a device to automatically record the bivariate relation between pressure of steam and its volume in a steam engine, with a view to calculating work done and improving efficiency.

A workable photographic process was first developed in 1827 by Joseph Niépce, and made practical with metal-coated glass plates by the Deguerre brothers in 1839. In 1878, Etienne-Jules Marey published his *La Méthode Graphique*, the first textbook on graphical methods. In this, he carefully reviewed the innovations by Playfair, Minard and others of static graphics, but he was more concerned with visualizing movement and change over time and space, in which both graphic recording and photography played key roles. He described a prodigious number of devices to record variation over time in physiological measures, such as the sphygmograph (pulse rate), cardiograph (heart rate), and polygraphs (galvanic skin response and other measures).

In other work, he developed a "photographic rifle" (following earlier experiments by Eadweard Muybridge) which allowed rapid photographs of objects (people, birds, etc.) in motion.[4] These images would contribute to the use in the Golden Age of what Tufte (1983a) later called "small multiples," the idea that change and variation could more readily be seen in multiple graphics arranged to allow easy comparison across a series.[5]

The wealth of data being collected in the early 19th century also created a need for serious number crunching. A large number of mechanical calculating devices were developed to meet this need, providing the rudiments of four-function calculators (addition, subtraction, multiplication, division).[6] In 1822, Charles Babbage [1791–1871] conceived of the Difference Engine, a mechanical device for calculating mathematical tables of logarithms and trigonometric functions and automatically printing the results, and later, the Analytical Engine (1837), a mechanical general-purpose programmable computer obtaining program instructions and data via punched cards (such as had been used on mechanical looms). Neither of these were actually constructed in his lifetime, but the idea of tabulating large volumes of data was in the air throughout the Golden Age.

For example, André-Michel Guerry (1864) collected massive amounts of data on crime and suicide in England and France over 25 years including 226,000 cases of personal crime classified by age, sex and other factors related to the accused and the crime, and 85,000 suicide records classified by motive; he invented an *ordonnateur statistique* to aid in the analysis and tabulation of these numbers. By 1890, in time for the decennial U.S. Census, Herman Hollerith [1860–1929] introduced a modern form of punched card to store numerical information, a keypunch device for entering data, and mechanical devices for counting and sorting the cards by columns of data.

## 2.5 Inventions in Statistical Graphics and Cartography

*Whatever relates to extent and quantity may be represented by geometrical figures. Statistical projections which speak to the senses without fatiguing the mind, possess the advantage of fixing the attention on a great number of important facts* —Alexander von Humboldt, 1811.

[4]It is said that Marey's "chronophotographs" provided an inspiration for Marcel Duchamp's *Nude descending a staircase*. See Braun (1992) for a richly illustrated account of Marey's work.

[5]The first instances I know of this technique are contained in Galileo's *Letters on Sunspots* and Christoph Scheiner's (1626) *Rosa Ursina sive Sol*, containing successive images of the position of sunspots across a series of days. Galileo drew these by hand, but Scheiner developed a "helioscope" and camera obscura to record these directly.

[6]The most extensive collection of such devices I know of is in the *Conservatoire des Artes et Métiers* in Paris.

2.5.1 *Statistical graphics.* In the first half of the 19th century, all of the modern forms of statistical graphics and thematic cartography were invented. Each contributed new elements to an emerging visual language designed to make quantitative data more comprehensible—to make the *relations* among numbers "speak to the eyes." The essential new idea here was that of graphic *comparison*—to show directly how one set of numbers compared to another over time or space by variation in visual attributes: length, position, angle, size, etc.

Line graphs were developed by William Playfair (1786, 1801) largely to show the changes in economic indicators (national debt, imports, exports) over time and to show the differences and relations among multiple time-series. Playfair also invented comparative bar charts to show relations of discrete series for which no time metric was available (e.g., imports from and exports to England) and pie charts and circle diagrams to show part-whole relations.

Playfair's initial uses of pie charts and circle diagrams (Playfair, 1801, Chart 2) actually introduced several novel principles for visual comparison simultaneously: (a) angular sectors to show part-whole, (b) diameter to show a total size, and (c) overlapping proportional circles to show relations among three entities, as in a Venn diagram (see Spence, 2005 for examples and discussion).

Over a few decades, a related idea arose, now called a "polar-area" diagram. If a pie chart is a plot in polar coordinates of $(r, \theta_i)$, with constant $r$ and varying $\theta_i$, the polar-area diagram is a plot of a cyclic phenomenon (e.g., deaths by month of year or day of week) of $(r_i, \theta)$, where the constant angle, $\theta$, divides the year or week into equal sectors, and the radius, $r_i$, is proportional to the $\sqrt{n_i}$, so the area of each sector represents the frequency or count, $n_i$.

The most famous example of this graphical method is the so-called "coxcomb diagram" (Figure 2) used by Florence Nightingale (1858) in her campaign to improve sanitary conditions for the British Army. The chart displays the causes of the deaths of soldiers during the Crimean war, divided into three categories: preventable infectious diseases (including cholera and dysentery, colored in blue), "wounds" (red) and "all other causes" (black). It showed immediately to the eye that most of the British soldiers who died during the Crimean War died of sickness rather than of wounds sustained in battle or other causes. The two separated parts compared death rates before and after March, 1855, when a delegation of Sanitary Commissioners was sent from London to improve hygiene in the camps and hospitals; again, the graphic comparison is direct and immediate: far fewer died after than before.

Nightingale was not the inventor of polar-area diagrams, even though this is almost always credited to her.[7] Rather, she was arguably the first to use this and other statistical graphs for political persuasion and popular impact—what Tukey (1990) later referred to as "interocularity": the graphic message hits you between the eyes.

In the Golden Age, polar-area diagrams would be put into service as an element of visual language in displays for more complex data. Some examples will appear in connection with the themes in Section 3.

2.5.2 *Thematic cartography.* During this same period, the first half of the 19th century, significant new forms of symbolism for thematic maps were introduced or extended. What is particularly interesting here is the diffusion of ideas between the emerging disciplines of statistical graphics and thematic cartography, where new techniques from each field influenced the other. For one thing, those who developed new methods for data-based maps were often not cartographers. For another, developers of data graphics often borrowed, and then extended, ideas from cartography; map-makers then returned the favor, extending those techniques in novel ways back to the geo-spatial context. Finally, in map-based graphics, the map was often secondary: used just as a background coordinate system, or even more striking—deformed to fit the data (see Figures 5 and 15).

One common theme was the desire to show intensity of a quantitative phenomenon or density of events in a geo-spatial context. As I noted earlier, the shaded choropleth map introduced by Dupin (1826) to display levels of education in France was quickly adapted by Guerry (1833, 1864) and others to explore relations among crime and other moral variables, but these represented essentially continuous phenomena with abrupt changes across geographical units. Montizon (1830) introduced a dot map, showing the population of France, with one dot proportional to a given population; this allowed for the representation of count data with more fine-grained variation. In the Golden Age, a dot map was the form chosen by John Snow (1855) to record the locations of deaths in the London cholera outbreak of 1854. These clustered mainly around the water pump on Broad Street, leading to his recognition of cholera as a water-borne disease.

[7] The first known use of this graphical method was by Guerry (1829a), which he called *courbes circulaires*, to show seasonal and daily variation in wind direction over the year and births and deaths by hour of the day. Slightly later, Léon Lalanne (1845) used a "windrose" diagram to show the frequency of wind directions continuously around the compass.

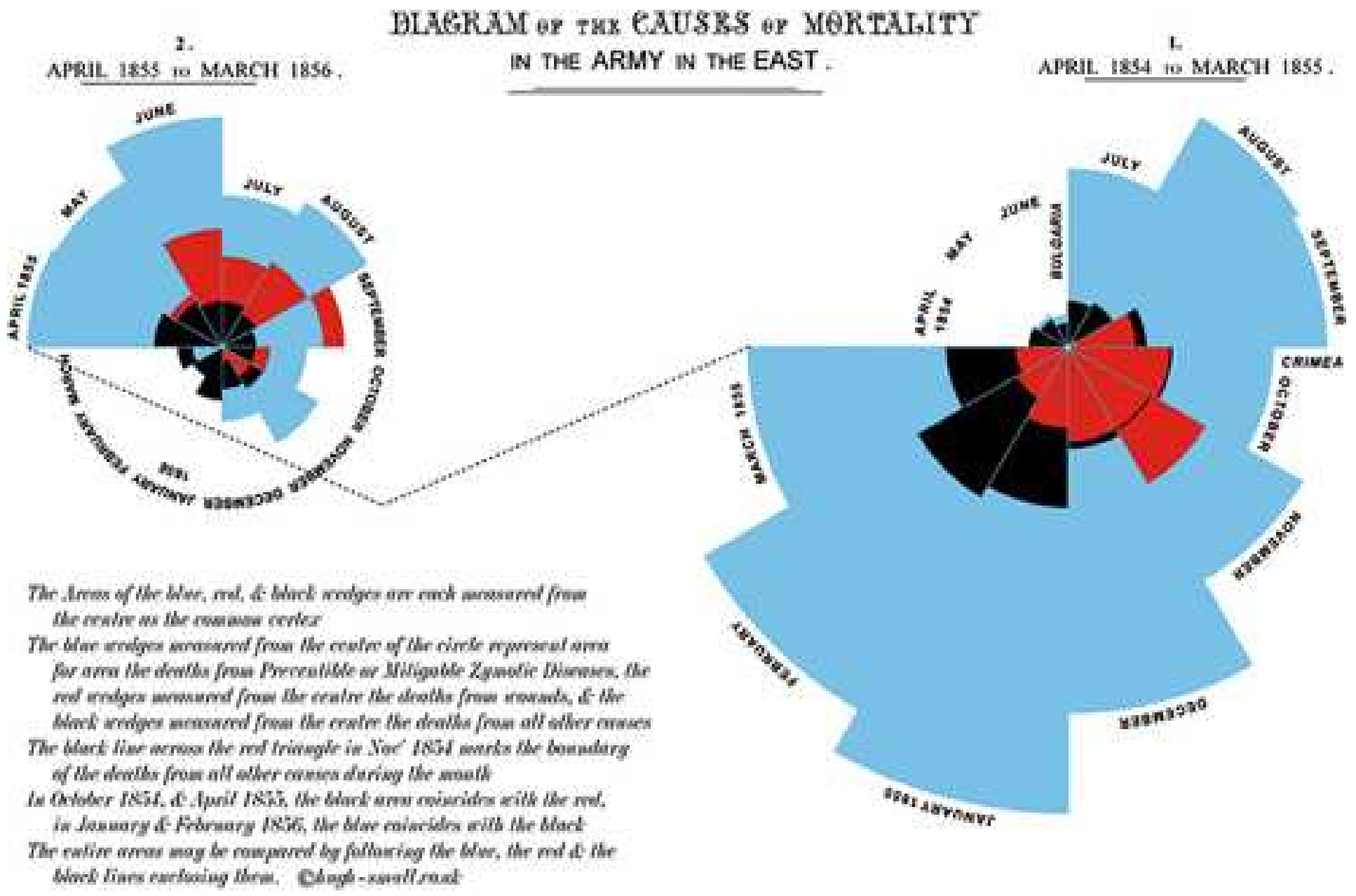

FIG. 2. *Multicomparative polar-area diagrams: Florence Nightingale, "Diagram of the causes of mortality in the army in the east." For each month, the number of deaths is shown by the area of sectors, distinguished as deaths from preventable diseases (blue), wounds (red) and other causes (black). Left/right: after and before March 1855. Source: Modern reproduction of the image from Nightingale (1858),* `http://hugh-small.co.uk/`.

A more abstract idea was that of showing isolines—contours of equal value—in a map-based representation of an essentially three-dimensional (3D) phenomenon. Edmund Halley's (1701) first thematic map showing isogons of constant magnetic declination set the stage for this technique. The concept was not further developed until 1817 when Alexander von Humboldt devised the first isotherm map (Wallis, 1973), showing lines of constant temperature in the Northern hemisphere.[8] Humboldt's "map" shows no geographical features other than a few place names; latitude and longitude are used simply to provide the coordinates for plotting the contours. Recognizing that average temperature depends strongly on altitude, he appended a contour plot of the relation between altitude and latitude.

It is important to highlight another important feature of these early contour maps. The data from which they were constructed were observed and recorded pointwise, at sparse, geographically scattered locations. The contours plotted by Halley, Humboldt and others were a further abstraction, based on *interpolation* and *smoothing*, by hand and eye, with the understanding that these phenomena must vary lawfully.[9]

The ideas of isolines and the contour map were then applied in a variety of other contexts, both data-based and map-based. The most extensive use appeared in Heinrich Berghaus's *Physikalischer Atlas* (Berghaus, 1838) (urged by von Humboldt who saw it as the graphic counterpart of his project describing the "Kosmos"). This broke no new ground in technique, but was the first big wholly thematic atlas, containing diagrammatic maps showing the distributions of climate, geology, plants, animals and so forth.

[8] See http://www.math.yorku.ca/SCS/Gallery/images/humboldt-isothermes1817.jpg.

[9] For example, Halley's map of isogons was based on about 150 directly recorded observations, collected over several years of voyages on the ship *Paramour*. The process of smoothing observational data graphically, to eliminate observational fluctuation and thereby reveal lawful regularity, was first described explicitly by Herschel (1833). Herschel regarded this as fundamental to scientific induction, analogous to Quetelet's use of the average to characterize a distribution.

New ground *was* broken by Léon Lalanne [1811–1892] (1845) with a general method for the graphic representation of tabular data of meteorological and other natural phenomena. An example is a contour plot of average soil temperature recorded for months of the year and hours of the day.[10] The $12 \times 24$ data table was complete and the phenomenon quite lawful, so interpolation and smoothing were straightforward. The result, however, enabled Lalanne to imagine the full 3D surface—he added front and side elevations showing the marginal contours of temperature by month and by hour, giving one of the first multiviews of complex bivariate data.

The final significant development in thematic cartography (for the present purposes) was a symbolism to show *movement* or change on a geographical background. In the 1830s, steam railways began to be developed rapidly in Britain and on the continent, with routes often determined solely by commercial interests. In a report for the Irish railway commission in 1836, the engineer Henry Drury Harness [1804–1883] (Harness, 1838; see also Robinson, 1955) pioneered the use of "flow lines," with width proportional to the quantity of interest to show the relative numbers of passengers or the total amount of traffic moving in a given direction. Harness's graphical invention had little impact initially. It was reinvented some years later, apparently independently, by both Alphonse Belpaire (in 1844) in Belgium and Charles Joseph Minard (1845) in France. In the hands of Minard and those who followed him, and with the advent of color printing, the flow map was developed to an art form, and provided some of the most potent graphical images of the Golden Age, as we shall see in Section 3.

2.5.3 *Nomograms and graphical calculation.* A final aspect of the graphic language that would contribute to the Golden Age, albeit indirectly, arose from the practical needs of civil and military engineers to provide easy means to perform complex calculations without access to anything more than a calculating diagram (or "nomogram"), a straightedge and pencil. For example, artillery and naval engineers created diagrams and graphical tables for calibrating the range of their guns. Léon Lalanne, an engineer at the Ponts et Chausées, created diagrams for calculating the smallest amount of earth that had to be moved when building railway lines in order to make the work time- and cost-efficient (Hankins, 1999).

Perhaps the most remarkable of these nomograms was Lalanne's (1844) "Universal calculator," which allowed graphic calculation of over 60 functions of arithmetic (log, square root), trigonometry (sine, cosine), geometry (area, circumference and surface of geometrical formas), conversion factors among units of measure and practical mechanics.[11] In effect, Lalanne had combined the use of parallel, nonlinear scales such as those found on a slide-rule (angles to sine and cosine) with a log–log grid on which any three-variable multiplicative relation could be represented by straight lines. For the engineer, it replaced books containing many tables of numerical values. For statistical graphics, it anticipated ideas of scales and linearization used today to simplify otherwise complex graphical displays.

I illustrate this slice of the Golden Age with Figure 3, a tour-de-force graphic by Charles Lallemand (1885) for precise determination of magnetic deviation ($\delta$) of the compass at sea in relation to latitude and longitude without calculation. This graphic combines many variables into a multifunction nomogram, where $\delta$ depends on seven values through complex trigonometric formulas. It incorporates 3D figures, juxtaposition of anamorphic maps, parallel coordinates and hexagonal grids. In using this device, the mariner projects his position at sea on the anamorphic map at the left through the upper central cone, then the grids and anamorphic maps at the right, and finally through the bottom central cone onto the scale of magnetic deviation. Voila! He can assure the crew they will be back home in time for Sunday dinner.

## 3. CONTRIBUTIONS TO THE GOLDEN AGE

It is not possible in this paper to describe more than a few contributions to the history of statistical graphics that exemplify the Golden Age. I have chosen three topics to illustrate general themes, not touched on above, that characterize the importance of the Golden Age. These concern: (a) the development of visual thinking, (b) the role of visualization in scientific discovery, and (c) graphical excellence, embodied in state statistical atlases.

[10] http://www.math.yorku.ca/SCS/Gallery/images/isis/fg14b.gif

[11] See `http://www.math.yorku.ca/SCS/Gallery/Lallane.jpg`.

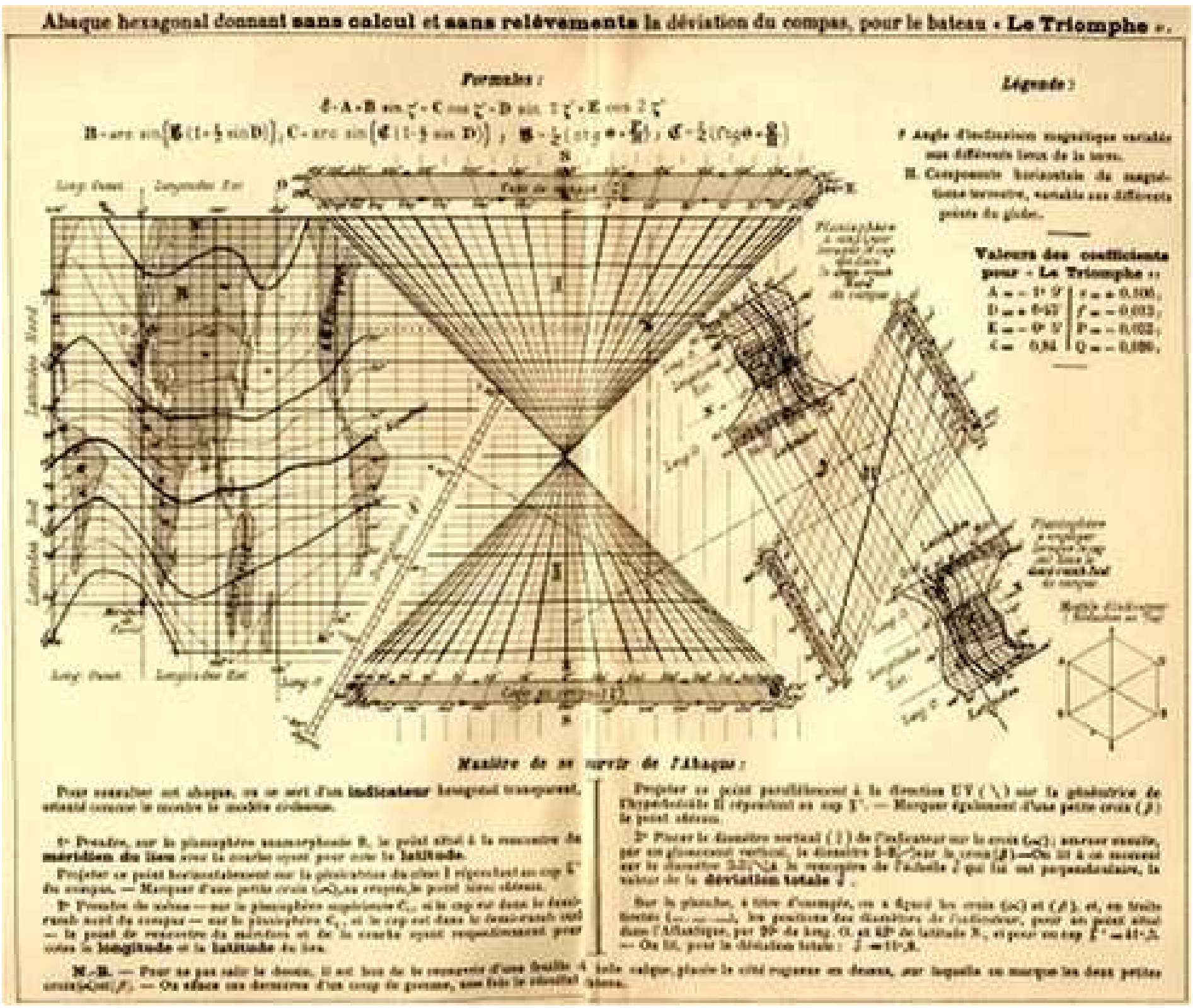

FIG. 3. *Multifunction nomogram combining diverse graphic forms: Charles Lallemand, "Hexagonal abacus giving the deviation of the compass without calculation, for the boat 'Triomphe'" (Abaque hexagonal donnant sans calcul et sans relévements la deviation du compass, pour le bateau 'Le Triomphe') (1885). Source: Ecole des Mines, Paris.*

## 3.1 The Graphic Vision of Charles Joseph Minard

*The dominant principle which characterizes my graphic tables and my figurative maps is to make immediately appreciable to the eye, as much as possible, the proportions of numeric results. . . . Not only do my maps speak, but even more, they count, they calculate by the eye.* —Minard (1862).

Until relatively recently, Charles Joseph Minard [1781–1870] was known only for one graphic work: his compelling portrayal of the terrible losses suffered by Napoleon's Grand Army in the 1812 campaign on Moscow. E. J. Marey (1878) said this graphic "defies the pen of the historian in its brutal eloquence," and 1983a called it "the best statistical graphic ever produced." I have reviewed Minard's graphical works elsewhere (Friendly, 2002b), but here I want to use him as an example, certainly not unique, of the rise of visual thinking and visual explanation that began in the early 19th century and came to fruition in the Golden Age.

Minard was trained as an engineer at the École Nationale des Ponts et Chausées (ENPC) and spent the rest of his life working there. However, a review of his work shows that he had two distinct careers there: In the first (1810–1842), he served as a civil engineer, designing plans for construction of canals and railways; in the second (1843–1869), he served as what can be called a visual engineer for the modern French state.

An example of visual thinking and visual explanation from his early career is shown in Figure 4. In 1840, Minard was sent to Bourg-Saint-Andèol to report on the collapse of a suspension bridge across the Rhône, constructed only 10 years before and therefore a major embarrassment for the ENPC. Minard's findings consisted essentially of this self-explaining before–after diagram. The visual message was immediate and transparent: apparently, the river bed beneath the supports on the up-stream side had eroded, leaving the bridge unsupported over a good portion of its width.

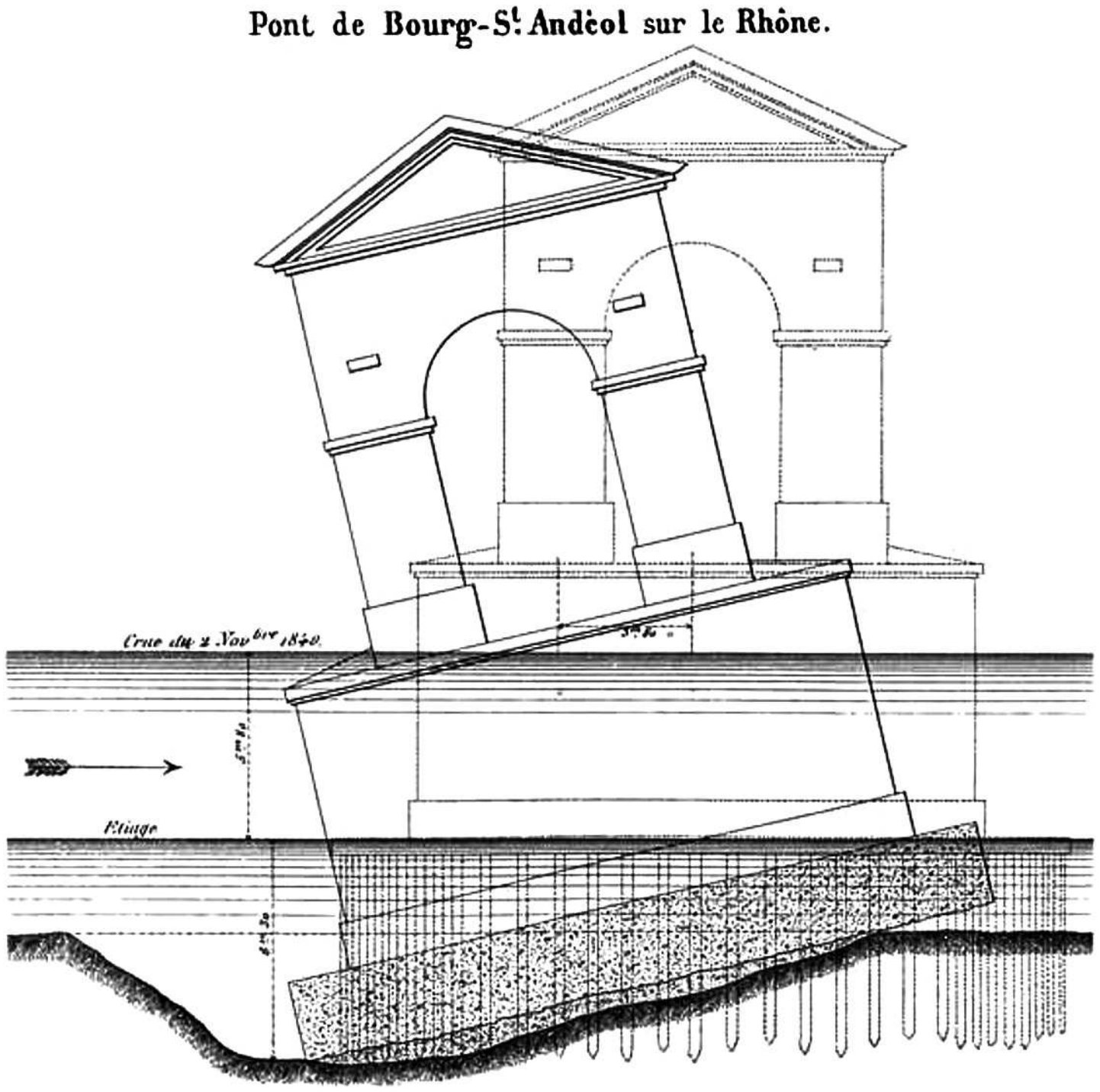

FIG. 4. *Engineering diagram: Why did the bridge collapse? A cross-sectional diagram showing one of the fixed bridge supports, providing a before–after comparison. Charles Joseph Minard (1840). Source: Tufte (1983a, 1983b).*

Minard produced 63 known graphic works[12] in the 1843–1869 period, which included *tableaux graphiques* (charts and statistical diagrams) and *cartes figuratives* (thematic maps). Before his retirement in 1851 his "bread and butter" topics concerned matters of trade, commerce and transportation: where to build railroads and canals? how to charge for transport of goods and passengers? how to visualize changes over time and differences over space? Most of his thematic maps were flow maps, which he developed to a near art form. His choice of the term *carte figurative* signals that the primary goal was to represent the data; the map was often secondary.

Perhaps the best illustration of these features of Minard's graphic vision is the pair of before–after flow maps shown in Figure 5. The goal here was to explain the effect that the U.S. Civil War had on trade in cotton between Europe and elsewhere. Again, the visual explanation is immediate and interocular: In 1858, most of the cotton imported to Europe came from the U.S. southern states—the wide blue swath that dominates the left figure. By 1862, the blockade of shipping to and from the South reduced this supply to a trickle, which came entirely through the port at New Orleans; some of the demand was met by Egyptian and Brazilian cotton, but the bulk of the replacement was imported from India. Note that, in order to accommodate the flow lines, he widened the English Channel and the Strait of Gibraltar, and reduced the coastline of North America to a mere cartoon form.

A final example of Minard's graphical vision is shown in Figure 6. In 1867 it was proposed that a new central post office be built in Paris; where should it be located? Minard's graphic approach was to draw the map with squares proportional to the population of each arrondisement. The obvious solution, to a visual engineer, was to build at the center of gravity of population density, shown by the small white dot inside the square on the right bank of the Seine.

### 3.2 Francis Galton's Graphic Discoveries

*When lists of observations are printed in line and column, they are in too crude a state for employment in weather investigations; after their contents have been sorted into Charts, it becomes possible to comprehend them; but it requires meterographic Maps to make their meaning apparent at a glance.* —Galton (1861), page 1.

[12] See http://www.math.yorku.ca/SCS/Gallery/minbib.html for a complete list.

My second theme in this section is the role that a recipe composed of richly recorded data, a spice rack containing graphic methods of many flavors, a pinch or two of technology, and, most of all, the hands of a visually masterful chef, would play in scientific discoveries in the Golden Age and into the early 20th century. There is no better exemplar than the graphic discoveries of Sir Francis Galton.

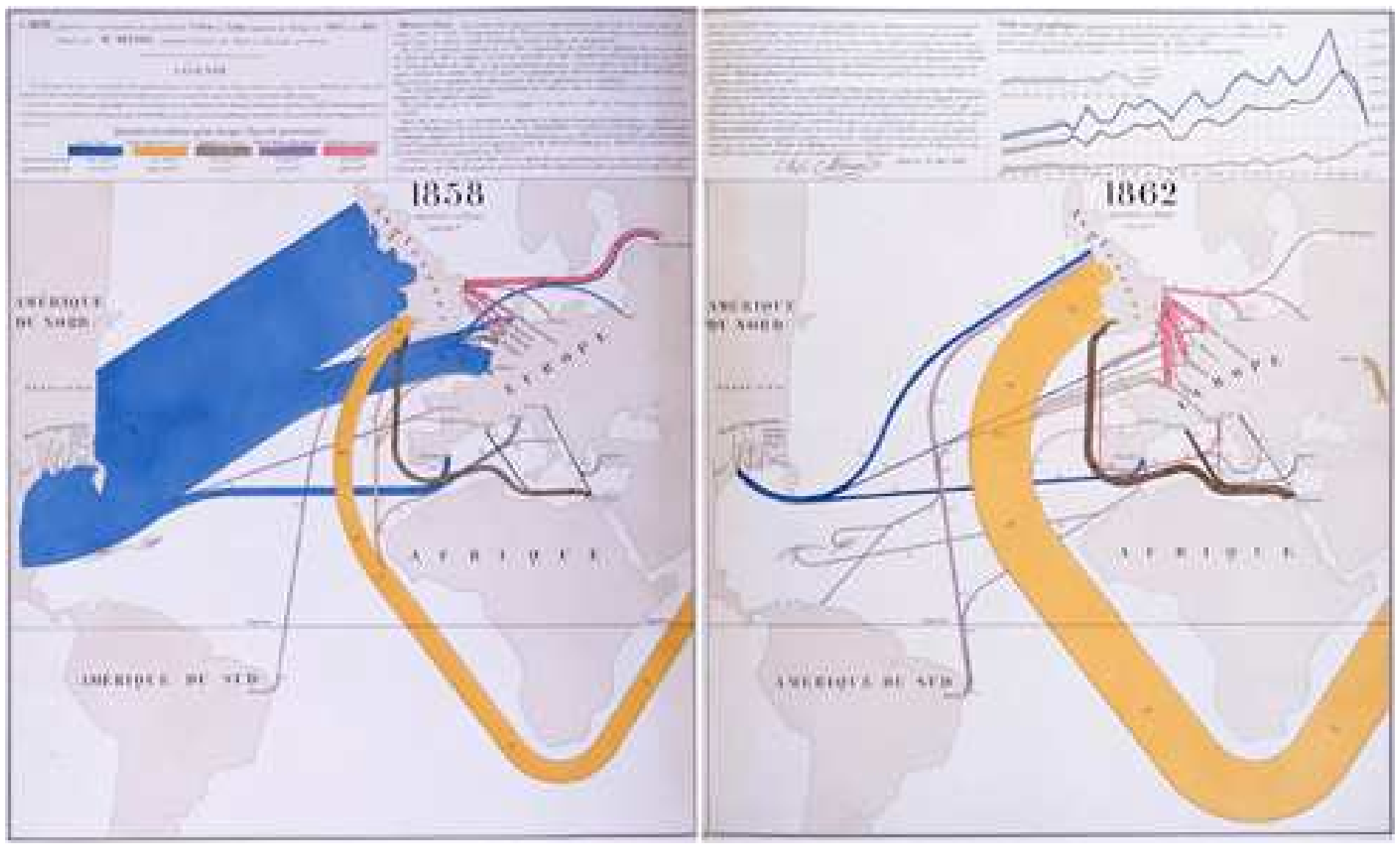

FIG. 5. *Comparative flow maps—effect of U.S. Civil War on trade in cotton. The import of raw cotton to Europe is shown from various sources to destination by colored flow bands of width proportional to the amount of cotton before (left: 1858) and after (right: 1862) the U.S. Civil War. Charles Joseph Minard, Carte figurative et approximative des quantitiés de coton en Europe en 1858 et 1862 (1863). Source: ENPC: Fol 10975.*

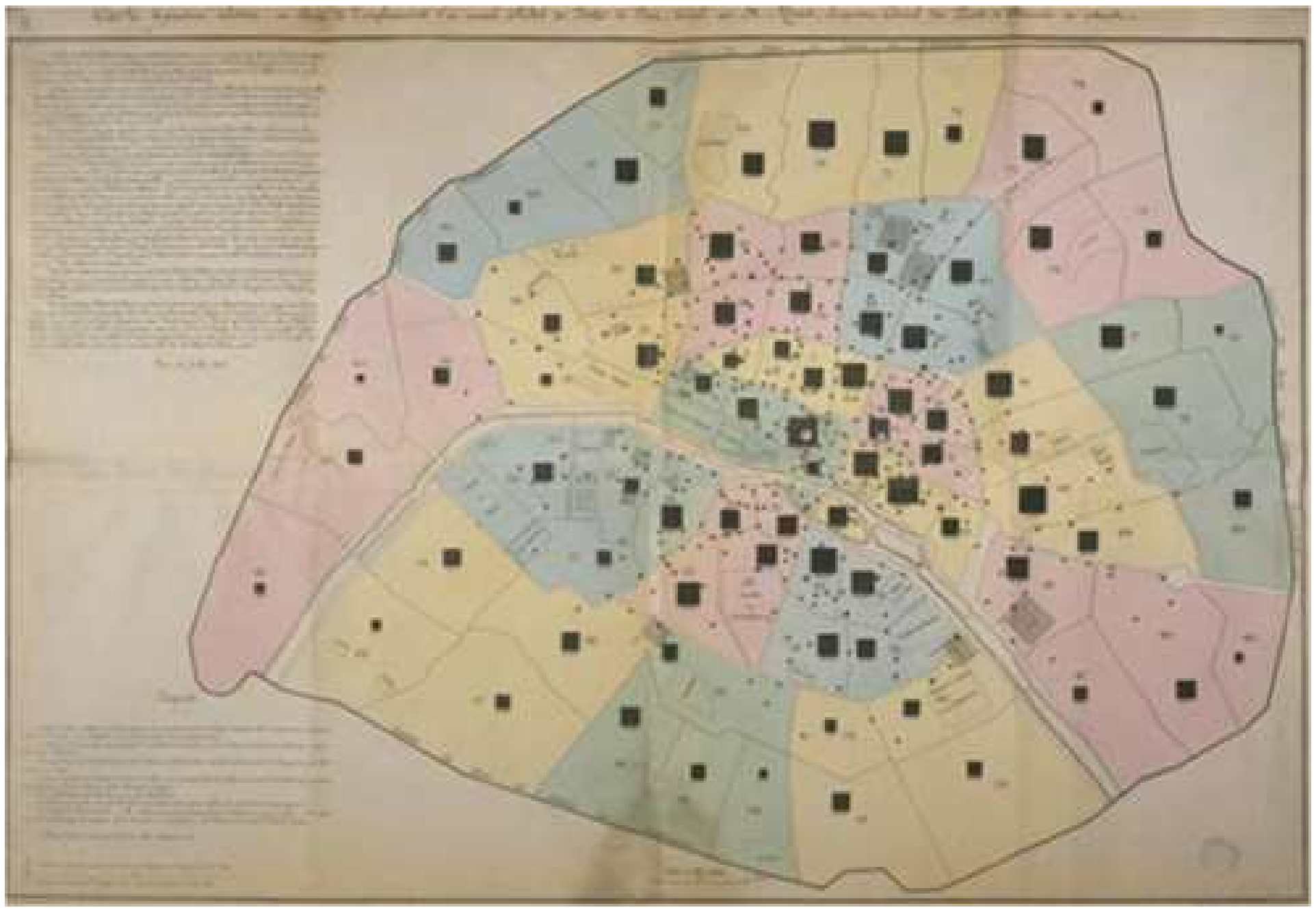

FIG. 6. *Center of gravity cartograph—where to build a new post office in Paris. The population of each arrondisement is shown at the administrative center of each arrondisement with a square proportional to population. The geographic center of gravity weighted by population is shown by the white dot. Charles Joseph Minard, Carte figurative relative au choix de l'emplacement d'un nouvel hôtel des postes de Paris (1865). Source: ENPC: Fol 10975, 10970/C589; BNF: Ge C 9553.*

Galton is most well known in statistics for his discovery of the characteristics of the bivariate normal frequency surface (Galton, 1886) that led to the invention of the method of regression and later to the theory of correlation by Karl Pearson (1896). In this, he profited greatly from the graphic tradition of isopleth lines—often smoothed and interpolated—on maps and contour diagrams of 3D relations that extended from Halley through Lalanne and others I have not mentioned.

Galton's statistical insight (Galton, 1886)—that, in a bivariate (normal) distribution (e.g., height of child against height of parent), (a) the isolines of equal frequency would appear as concentric ellipses, and (b) the locus of the (regression) lines of means of $y \mid x$ and of $x \mid y$ were the conjugate diameters of these ellipses—was based largely on visual analysis from the application of smoothing to his data. As Pearson would later say, "that Galton should have evolved all this from his observations is to my mind one of the most noteworthy scientific discoveries arising from pure analysis of observations" (Pearson, 1920, page 37). This was only one of Galton's discoveries based on graphical methods in general and visual smoothing in particular. See Friendly and Denis (2005) for a modern reanalysis of the visual ideas here.

I am going out of order in time, but I want to suggest that Galton had achieved an even more notable graphic discovery 25 years earlier, in 1863—the relation between barometric pressure and wind direction that now forms the basis of modern weather maps. It is not too far a stretch to claim this as a best example of a scientific discovery achieved almost entirely through graphical means, "something that was totally unexpected, and purely the product of his high-dimensional graphs" (Stephen Stigler, personal communication, quoted by Wainer, 2005).

Galton, a true polymath, began an interest in meteorology in about 1858, after he was appointed a director of the observatory at Kew. Many scientific questions related to geodesy, astronomy and meteorology were suggested to him by this work, but in his mind any answers depended first on systematic and reliable data, and second on the ability to find coherent patterns in the data that could contribute to a general understanding of the forces at play.

In 1861 he began a campaign to gather weather data from weather stations, lighthouses and observatories across Europe, enlisting the aid of over 300 observers. His instructions included a data collection form (Figure 7) to be filled out thrice daily, for the entire month of December, 1861, with barometric pressure, temperature, wind direction and speed, and so forth to be entered in the form. From the returns, he began a process of graphical abstraction, eventually published as *Meterographica* (Galton, 1863b). Altogether, he made over 600 maps and diagrams, using lithography and photography in the process.

In the first stage, he constructed 93 maps (three per day, for each of 31 days) on which he recorded multivariate glyphs using stamps or templates he had devised to show rain, cloud cover, and the direction and force of the wind, as shown in Figure 8. He explained that it was just as precise to use these symbols as the letters N, NNW, NW, etc. to express wind direction, but the icons "have the advantage of telling their tale directly to the eye."

Together with these he made iconic maps of barometric pressure (Figure 9). He says "these require a similar pictorial treatment to that employed for geographical elevations; in other words, areas of elevation and depression of the barometer &c, must be pictured by contour lines and shadings, on the same principles as mountains and valleys." In these he used classed symbols, with a bipolar scale of direction and intensity of the deviation from an average, though he says "they seem to me to be clear so far as they go, though the curves of nature are represented at a serious disadvantage by the mosaic work of types."[13]

From these 93 glyph maps and iconic 3D maps, Galton noticed something striking. At this time there was a theory of cyclones, suggesting that in an area of low barometric pressure, winds spiraled inward, rotating counterclockwise. Galton was able to confirm this from his charts, but noticed something else, which provided for a more general theory of weather patterns: across geographic space, areas of

[13] In these barometric maps, I would have been more pleased if Galton had published graphical versions using contour lines, as he suggested, rather than the admittedly inferior discrete representations. There is no indication in his meteorological writings if he ever drew the daily isobar versions, or how he passed from iconic maps to the small-scale summaries shown in Figure 10. He says that the original and intermediate versions of many of his charts and maps were deposited with the British Meteorological Society, but I have not been able to locate these so far.

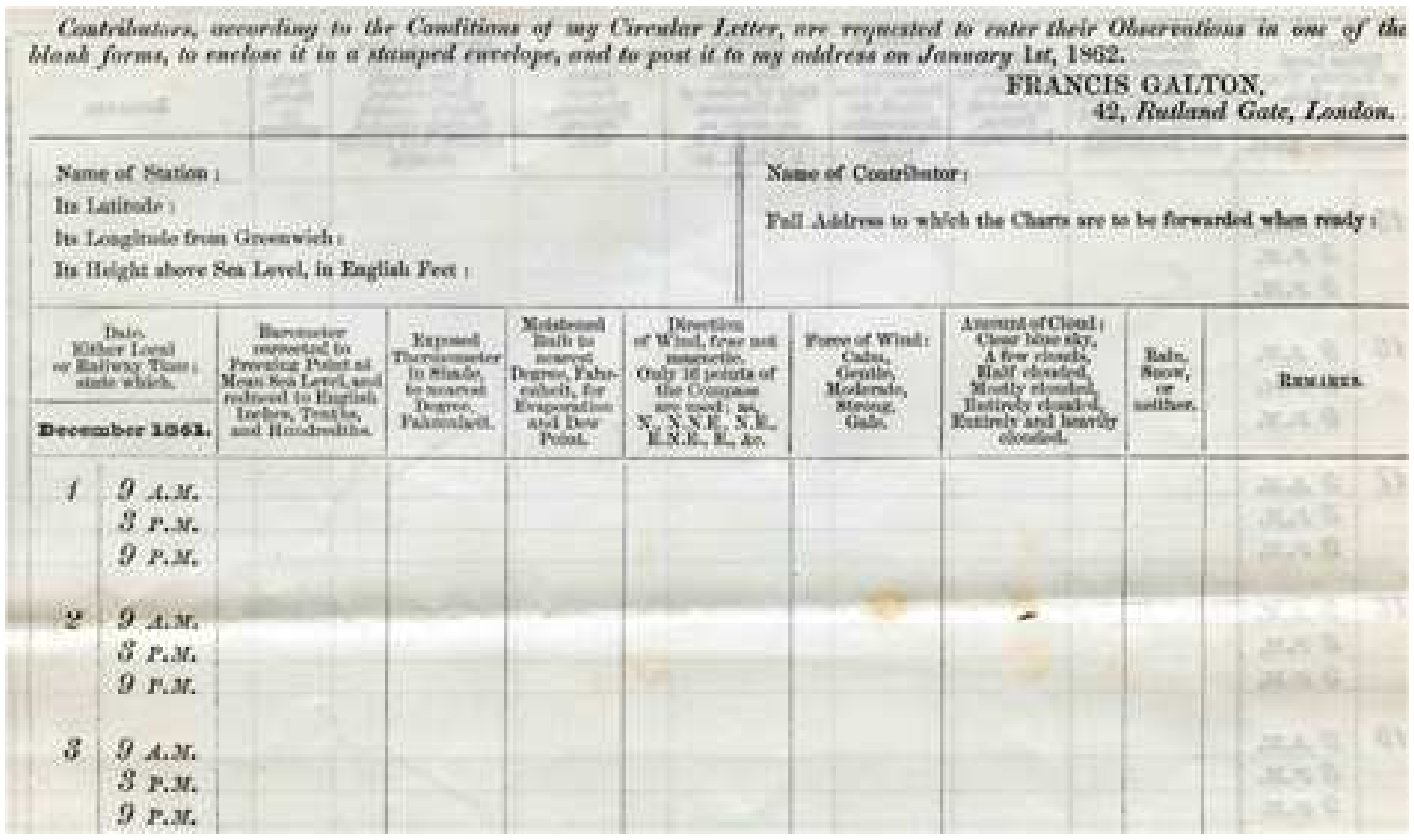

*Contributors, according to the Conditions of my Circular Letter, are requested to enter their Observations in one of the blank forms, to enclose it in a stamped envelope, and to post it to my address on January 1st, 1862.*

FRANCIS GALTON,
*42, Rutland Gate, London.*

Name of Station :
Its Latitude :
Its Longitude from Greenwich :
Its Height above Sea Level, in English Feet :

Name of Contributor :
Full Address to which the Charts are to be forwarded when ready :

| Date. Either Local or Railway Time; state which. December 1861. | | Barometer corrected to Freezing Point at Mean Sea Level, and reduced to English Inches, Tenths, and Hundredths. | Exposed Thermometer in Shade, to nearest Degree, Fahrenheit. | Moistened Bulb to nearest Degree, Fahrenheit, for Evaporation and Dew Point. | Direction of Wind, true not magnetic. Only 16 points of the Compass are used; as, N., N.N.E., N.E., E.N.E., E., &c. | Force of Wind: Calm, Gentle, Moderate, Strong, Gale. | Amount of Cloud: Clear blue sky, A few clouds, Half clouded, Mostly clouded, Entirely clouded, Entirely and heavily clouded. | Rain, Snow, or neither. | Remarks. |
|---|---|---|---|---|---|---|---|---|---|
| *1* | *9* A.M. | | | | | | | | |
| | *3* P.M. | | | | | | | | |
| | *9* P.M. | | | | | | | | |
| *2* | *9* A.M. | | | | | | | | |
| | *3* P.M. | | | | | | | | |
| | *9* P.M. | | | | | | | | |
| *3* | *9* A.M. | | | | | | | | |
| | *3* P.M. | | | | | | | | |
| | *9* P.M. | | | | | | | | |

FIG. 7. *Data collection form. Top portion of Galton's form sent to observers to record weather variables throughout the month of December, 1861. Source: Galton (1863b), private collection.*

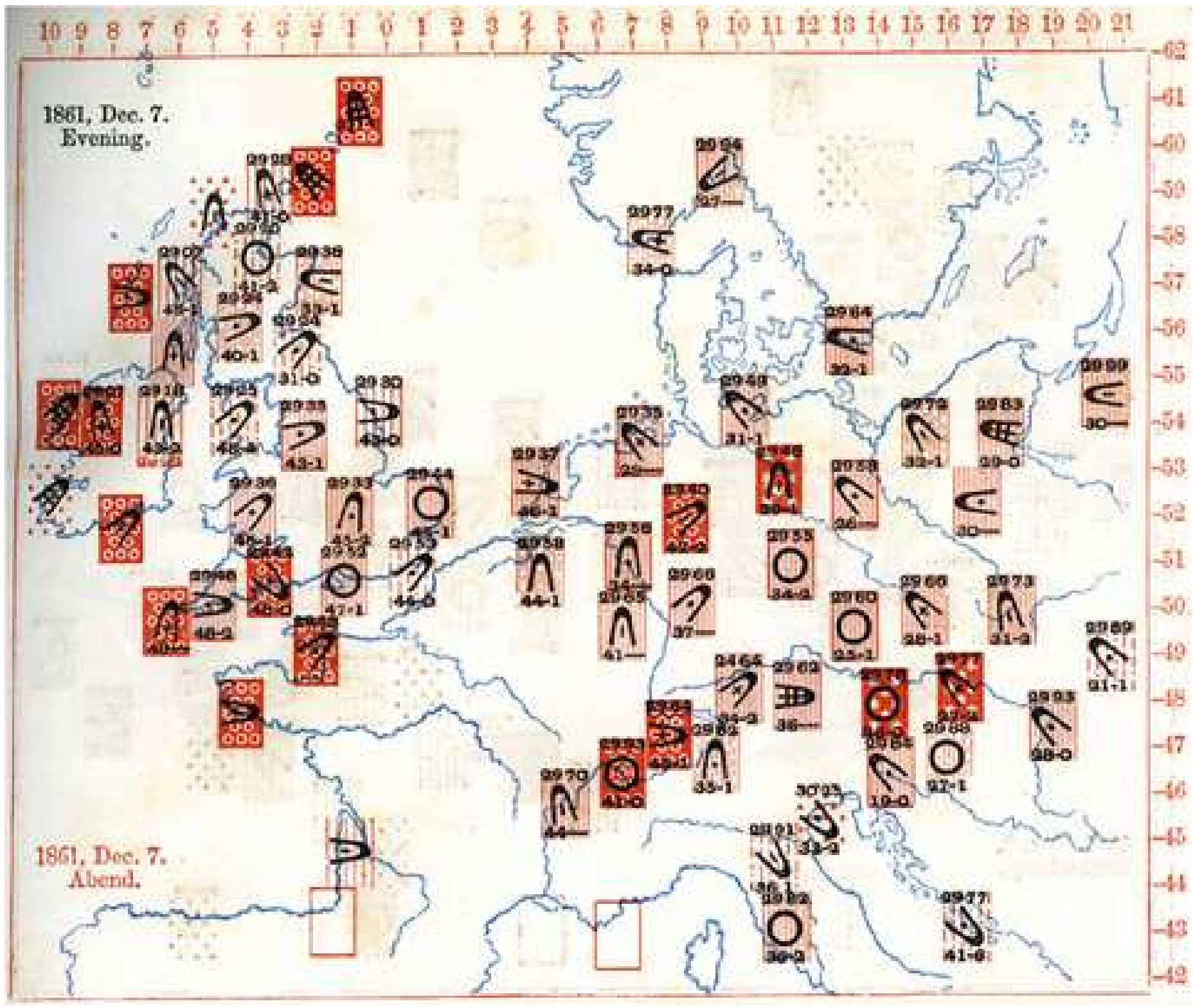

FIG. 8. *Multivariate glyph map. Galton's glyph map of wind, cloud cover and rain on the evening of December 6, 1861. The U-shaped icon opens toward the direction of the wind and is filled in relation to its strength; a circle indicates calm. Stippled and hatched backgrounds range from clear, through degrees of cloud to snow and rain. Source: Galton (1863b), private collection.*

high barometric pressure also corresponded to outward spiral wind in clockwise direction. He termed this relation an "anticyclone" (Galton, 1863a).

What was key to confirming these observations was his ability to see *relations* of wind direction and pressure over space and *changes* of these over time. In a second stage of abstraction, he reduced the data for each day to a $3 \times 3$ grid of miniature abstract contour maps, showing in the rows barometric pressure, wind direction and rain, and temperature, with columns for morning, noon and afternoon. In these he used color, shading and contours to show approximate isolevels and boundaries, and arrows to show wind direction. He assembled these for all 31 days into a single two-page chart of multivariate schematic micromaps, of which the right-hand page is shown in Figure 10; the legend for the symbols appears in Figure 11. Conveniently, it turned out that barometric pressure was generally low in the first half of December and high in the second half. The correlated directions of the arrows for wind direction confirmed the theory. He explained these results with reference to Dove's Law of Gyration (Galton, 1863a). A prediction from this and Galton's cyclone–anticyclone theory, that a reversed pattern of flow should occur in the southern hemisphere, was later confirmed.

Galton's discovery of weather patterns illustrates the combination of data, visual thinking and considerable labor to produce a theoretical description. His further work in meteorology also illustrates the translation of theory into practical application, another feature we find in the Golden Age. From 1861–1877, he published 17 articles dealing with meteorological topics, such as how charts of wind direction and intensity could be translated into charts of travel time for mariners (Galton, 1866). On April 1, 1875, the London *Times* published a weather chart prepared by Galton, the first instance of the modern weather maps we see today in newspapers worldwide.[14]

## 3.3 Statistical Atlases

*Let these facts be expressed not alone in figures, but graphically, by means of maps and diagrams, appealing to a quick sense of form and color and "clothing the dry bones of statistics in flesh and blood," and their study becomes a delight rather than a task.*

—Henry Gannett, preface to the *Statistical Atlas of the Tenth Census*, 1880.

My final theme concerns the graphical excellence that I take as a primary characteristic of the Golden Age. As noted earlier, the collection, organization and dissemination of official government statistics on population, trade and commerce, social and political issues became widespread in most of the European countries from about 1820 to 1870. In the United States, the first official census was carried out in 1790 to provide population data for proportional representation in the House of Representatives required by the U.S. Constitution.[15] By 1850, the need to provide a statistical summary of the nation and plan for future growth impelled the Census Bureau to gather data on a greatly expanded list of topics, including aspects of manufacturing and resources, taxation, poverty and crime. However, for the most part, these early national summaries of official statistics were presented in tabular form.

In the period from about 1870 the enthusiasm for graphic representation took hold in many of the state statistical bureaus in Europe and the United States, resulting in the preparation of a large number of statistical atlases and albums.[16] As befits state agencies, the statistical content and presentation goals varied widely, and the subject matter was often mundane. The resulting publications, however, are impressive, for their wide range of graphic methods and often for the great skill of visual design they reflect. As we shall see, they often anticipated graphical forms and ideas that were only re-invented in the period from about 1970 to the present.

3.3.1 *L'Album de Statistique Graphique.* Minard's graphic works at the ENPC were very influential

[14]See Monmonier (1999) for a cartographic perspective on the history of weather maps.

[15]This first U.S. Census, under the responsibility of Thomas Jefferson, Secretary of State, was carried out by U.S. Marshals on horseback. They counted 3.9 million inhabitants.

[16]Although these were separate works, it is worth noting that the central characters in France (Émile Levasseur, Émile Cheysson), Germany (Georg von Mayr, Hermann Schwabe), England (William Farr, Joseph Fletcher), the United States (Francis Walker, Henry Gannett) and elsewhere were well acquainted with each other's work, from international expositions, conferences, and informal exchanges. Among these, the International Statistical Congresses, organized in 1857 by Quetelet and others, soon became an international forum for debating the use of graphical methods and attempts to develop international standards; see, for example, International Statistical Congress (1858), Schwabe (1872), Palsky (1999).

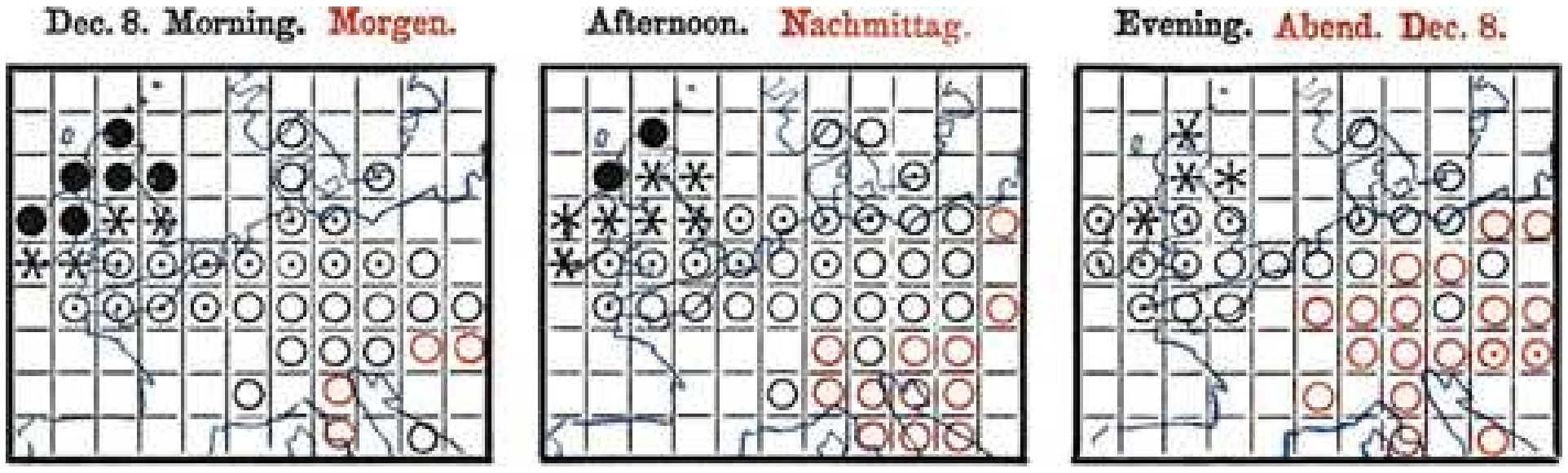

FIG. 9. *Iconic 3D barometric maps, bipolar scale. Galton's barometric maps for December 6, 1861. Red and black symbols represent respectively lower and higher barometric pressure than average, with degrees of divergence ranging from ○ through ⊙, ∗ to ". Source: Galton (1863a, 1863b), private collection.*

in government bureaus in France, so much so that nearly every minister in the Ministry of Public Works from 1850 to 1860 had his official portrait painted with one of Minard's works in the background (Chevallier, 1871, page 17). In March 1878, a bureau of statistical graphics was established in this ministry under the direction of Émile Cheysson [1836–1910], who was also an engineer at the ENPC, and who had played a major part in the committees on the standardization of graphical methods at the International Statistical Congresses from 1872 on.

By July 1878, the new bureau was given its marching orders and charged to "prepare (*figurative*) maps and diagrams expressing in graphic form statistical documents relating to the flow of passenger travel and freight on lines of communication of any kind and at the seaports, and to the construction and exploitation of these lines and ports; in sum, all the economic facts, technical or financial, which relate to statistics and may be of interest to the administration of public works" (Faure, 1918, page 294; Palsky, 1996, pages 141–142).

From 1879 to 1897 the statistical bureau published the *Album de Statistique Graphique* (Ministère des travaux publics, 1897).[17] They were published as large-format quarto books (about 11 × 15 in), and many of the plates folded out to four or six times that size, all printed in color and with great attention to layout and composition. The number of plates varied from 12 (in 1879) to 34 (in 1886), with most containing 21–23. I concur with Funkhouser (1937), page 336, that "the *Albums* present the finest specimens of French graphic work in the century and considerable pride was taken in them by the French people, statisticians and laymen alike." These volumes can be considered the pinnacle of the Golden Age, an exquisite sampler of nearly all known graphical forms, and a few that make their first appearance in these volumes.

Cheysson, in accord with his directions, designed the *Albums* to portray two series of themes or topics: (a) the recurrent one, concerning economic and financial data related to the planning, development and administration of public works—transport of passengers and freight, by rail, on inland waterways and through seaports, the bread and butter of their mandate; (b) occasional topics, that varied from year to year, and that included such subjects as agriculture, population growth, transport and international expositions in Paris, and so forth.

The greatest proportion of the plates in the *Albums* are cartograms, of one form or another; many of these use the form of the flow map pioneered by Minard, but extended by Cheysson to show more complex data. For example, there are quite a few plates using double- or multiflow lines to show two or more aspects of movement simultaneously (e.g., revenue from passengers and freight). Circular and radial forms also proved popular, and included pie maps, and diagrams or maps using stars, radial axes, polar time-series, and proportional circles as visual symbols for quantities shown. Some special and novel graphic forms included maps with mosaic displays [multiply divided squares; see Friendly (2002a) for a history of this graphic form], anamorphic maps and "planetary" diagrams. Basic chart forms were also used—line, bar and time-series graphs, sometimes embedded in another graphic or map to show some aspect of the data not represented in the principal display.

[17]Only two complete collections of these volumes are known to exist: one in the Tolbiac branch of the Bibliothèque Nationale de France in Paris, the other held collectively by *les chevaliers des Album de Statistique Graphique.*

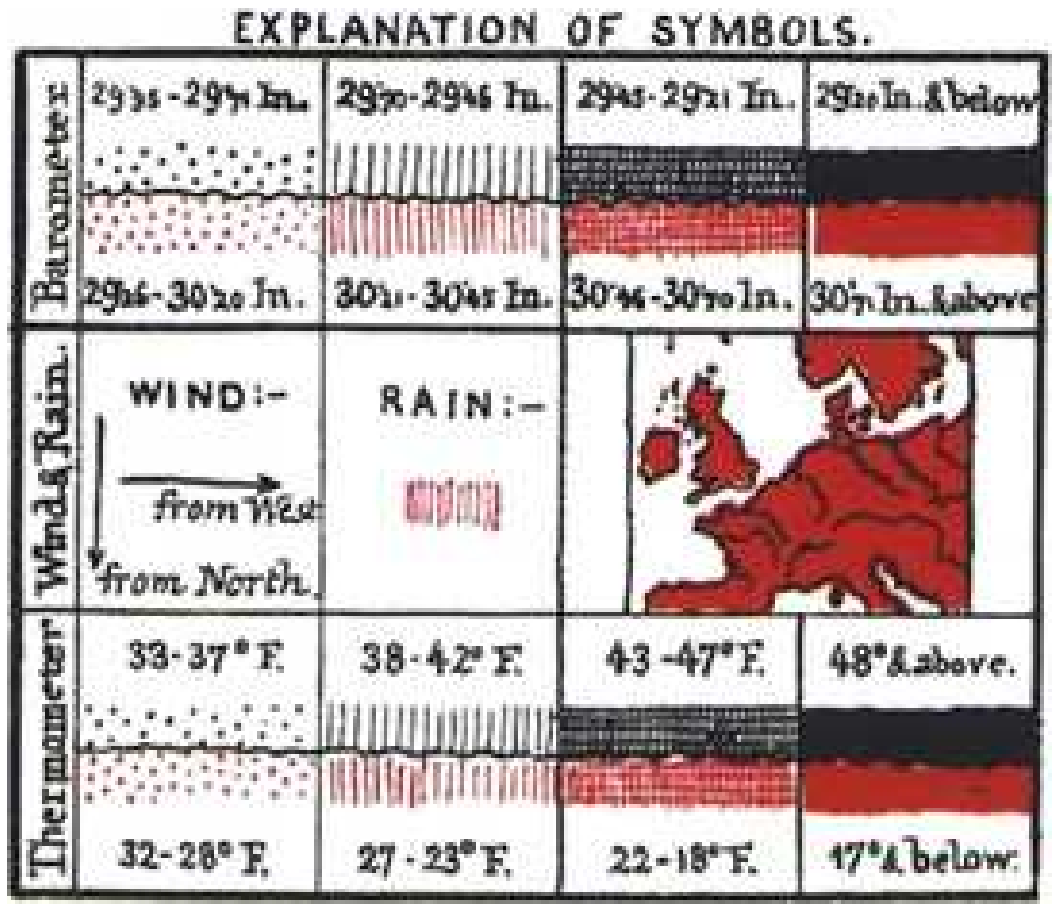

FIG. 11. *Multivariate schematic micromaps: Legend. Legend in the bottom right of Figure 10, showing the use of color, shape, texture and other visual attributes to portray quantitative variables. Source: Galton (1863b).*

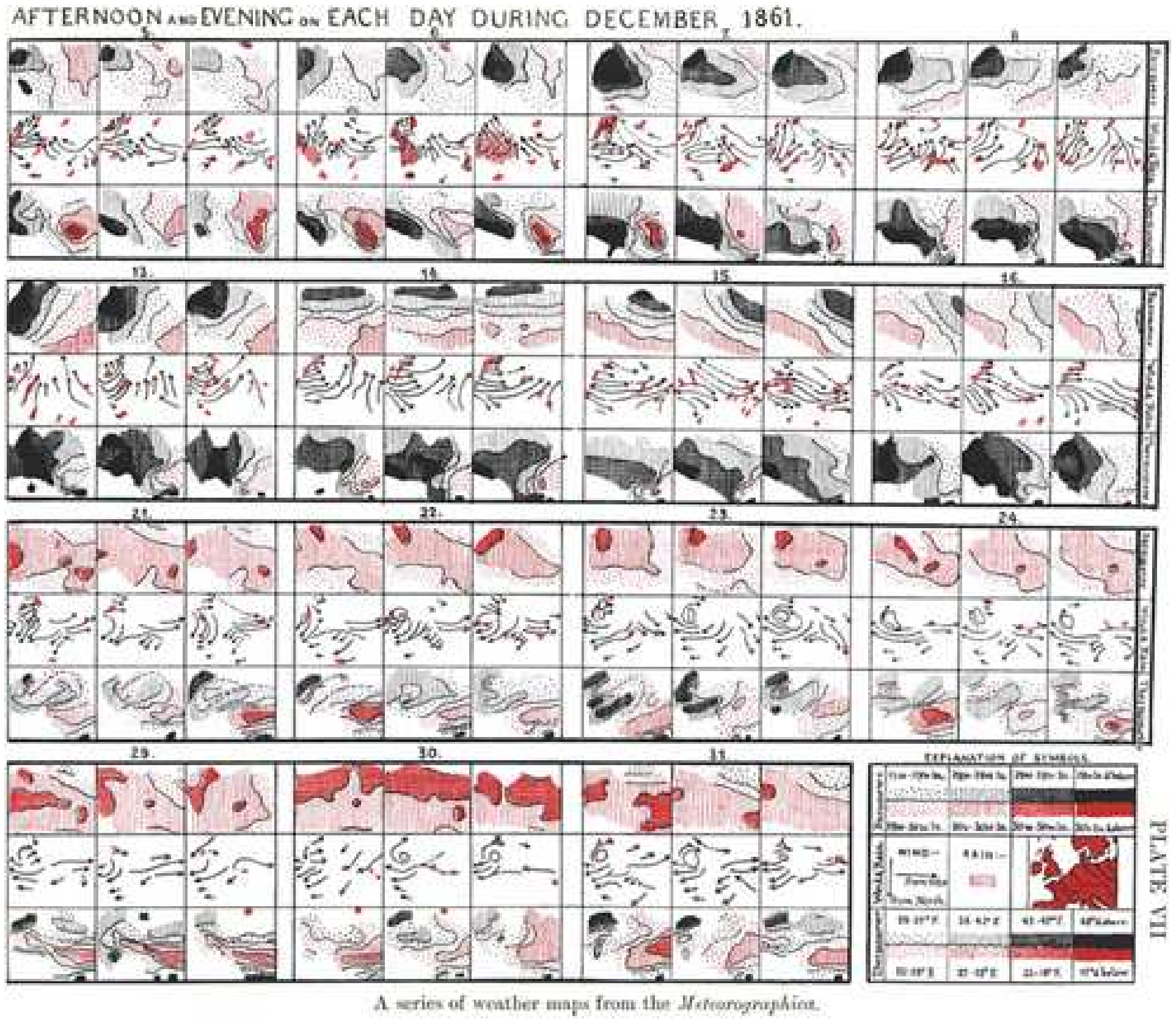

FIG. 10. *Multivariate schematic micromaps. Francis Galton, "Charts of the Thermometer, Wind, Rain and Barometer on the Morning, Afternoon and Evening on Each Day during December 1861." Each daily panel is a* $3 \times 3$ *display of the combinations of barometric pressure, wind and rain and temperature by morning, noon and afternoon. Source: Galton (1863b), Pearson (1930), private collection.*

The selection of images from the *Albums* described below is not representative; for that, see Palsky (1996), pages 141–161 and Wainer (2003) for two more examples. Instead, I focus on the images I consider most notable and imaginative; these are mostly taken from the occasional, special topics of the *Albums*.

Demographic data were not often presented in the *Albums*, but Figure 12 from the 1881 album provides a particularly attractive graphic design. It uses polar time-series diagrams on a map of France to show the changes in population by department at the 5-year intervals of the national census from 1801 to 1881. For each department, the central white circle shows the value in 1841, the middle of the series; the sectors are shaded red for years where the population was less than that value and are shaded blue–green for years where the population exceeded the 1841 value. The departments ceded to Germany in 1871 are handled differently and shaded in black. In addition to being visually pleasing, the scaling relative to the central value for 1841 provided a simple way to compare the trends across departments.

The 1889 volume followed the universal exposition in Paris of that year, and several novel graphic designs were used to provide an analysis of data related to this topic. Figure 13 uses what are now called *star* or *radar* diagrams to show attendance at each of the universal expositions held in Paris: 1867, 1878 and 1889. These are laid out as a two-way table of plots, in a form we now call a "trellis display," designed to allow comparisons of the rows (years) and columns (months). Each star diagram shows daily attendance by the length of the ray, in yellow for paid entrance and black for free admissions, with Sundays oriented at the compass points. In this display, we can see: (a) attendance increased greatly from 1867 to 1889; (b) Sundays were usually most well-attended; and (c) in 1889, there were a number of additional spikes, mostly holidays and festivals, that are noted on the graphs with textual descriptions.

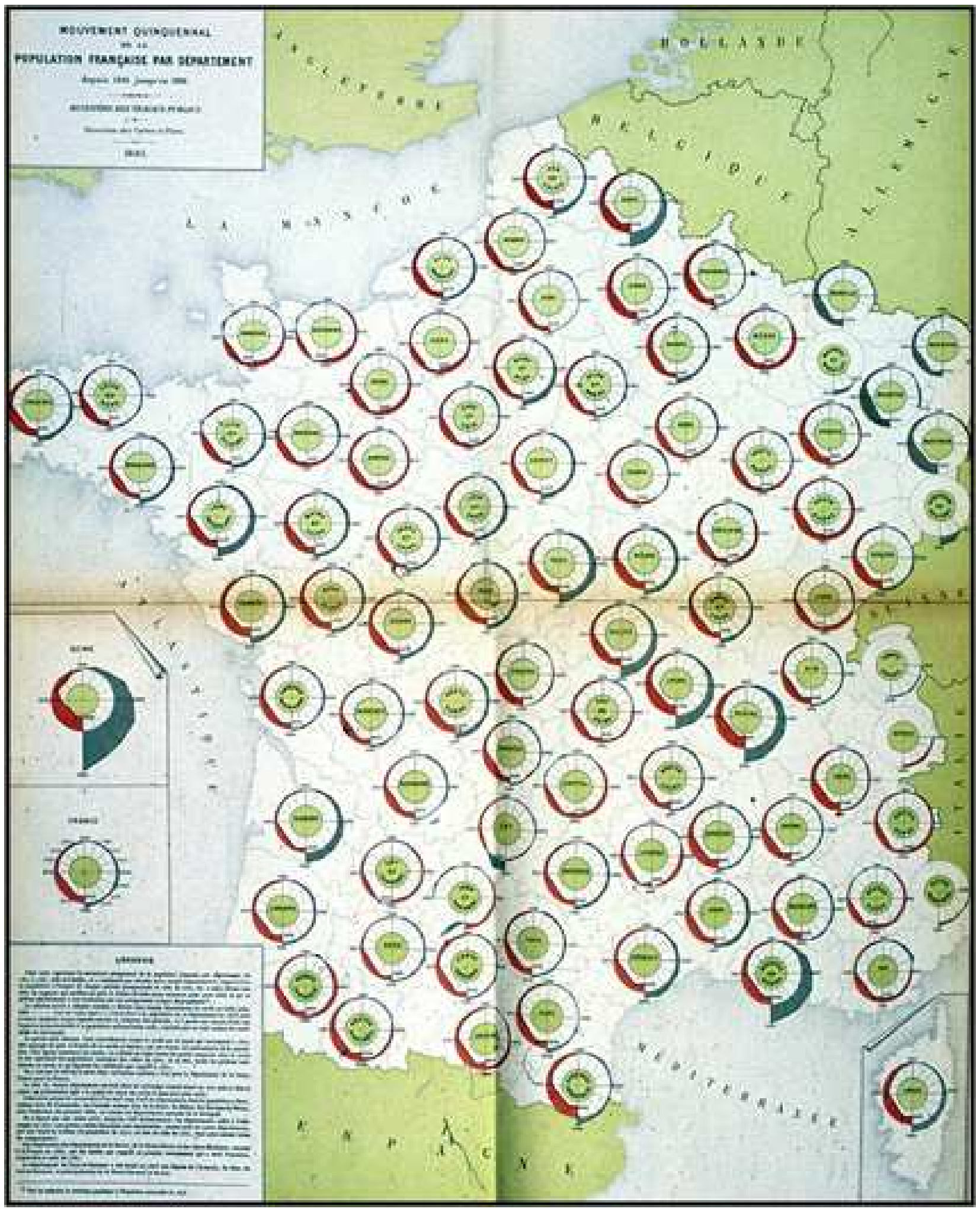

FIG. 12. *Spiral time-series diagram. "Five-year changes in the population by department from 1801 to 1881" (Mouvement quinquenial de la population par départment depuis 1801 jusqu'en 1881). The graphic design was intended to highlight the relative changes in each department over time. Source: Album de Statistique Graphique, 1884, Plate 25.*

What effect did these expositions have on attendance in theaters? Figure 14 shows polar diagrams with the area of each sector proportional to gross receipts for the major theaters in Paris in the years from 1878 to 1889. The expo years of 1878 and 1889 are shaded yellow, while others are shaded red. To aid interpretation according to Paris geography, these figures are placed on a map of the right bank of Paris, with theaters off-scale shown in boxes.

Figure 15 is yet another singular plate, this from the 1888 *Album*. It uses what is now called an *anamorphic map* to show how travel time in France (from Paris) had decreased over 200 years, and does this simply by shrinking the map. Here, the outer out-line of the map represents, along each radial line, the travel time to various cities in 1650. These lines are scaled in proportion to the reduced travel time in the years 1789, 1814, ... , 1887, with the numerical values shown in the table at the bottom right and along each radial line.[18] The outline of the map of France was then scaled proportionally along those radial lines. What becomes immediately obvious is that the shrinking of travel times was not uniform;

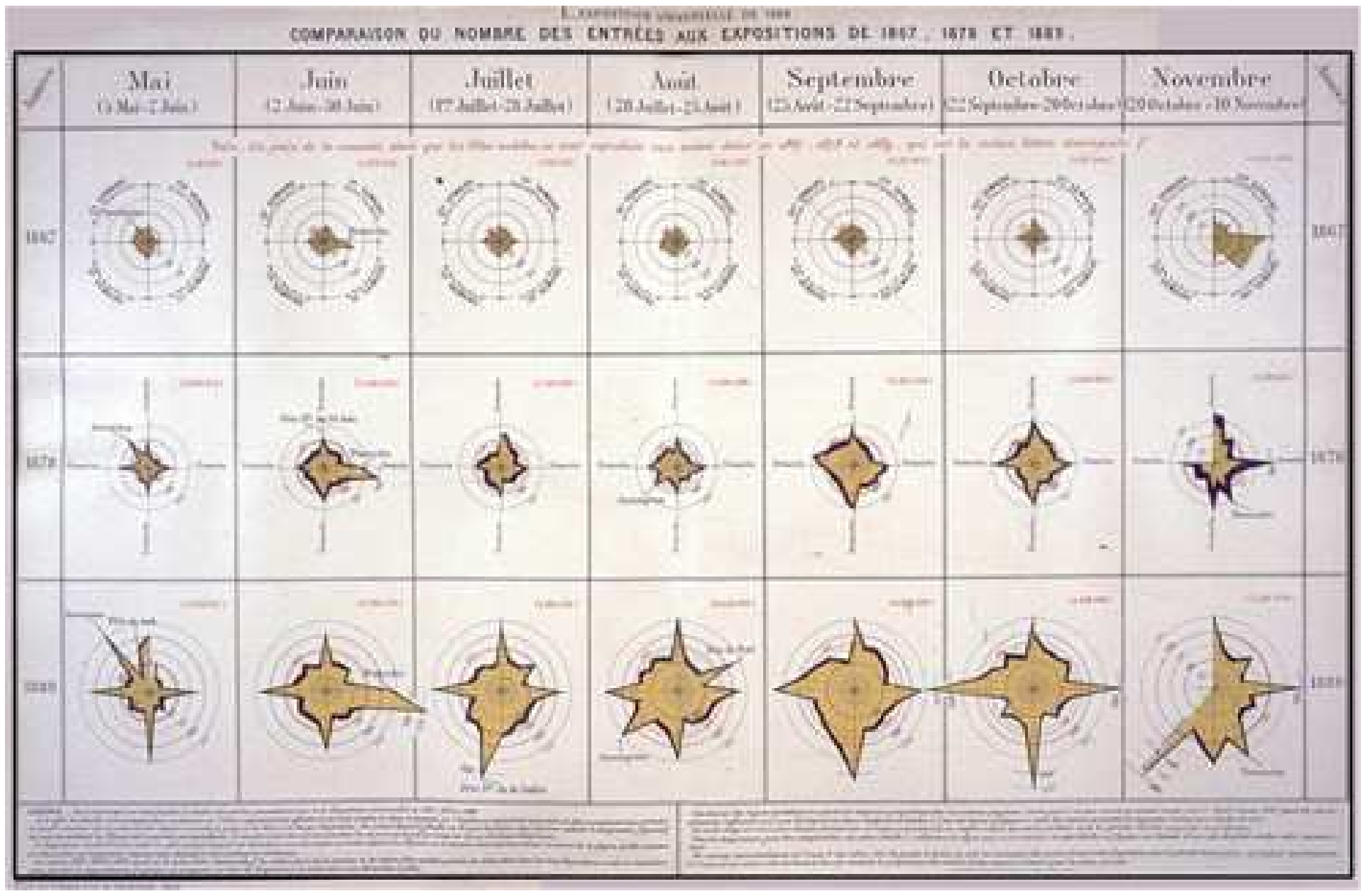

FIG. 13. *Two-way star/radar diagrams. "Comparison of the Numbers Attending the Expositions of 1867, 1878 and 1889" (Exposition Universelle de 1889: Comparaison du Nombre des Entrées aux Expositions de 1867, 1878 et 1889). Each star-shaped figure shows the number of paid entrants on each day of the month by the length of the radial dimension. Source: Album de Statistique Graphique, 1889, Plate 21.*

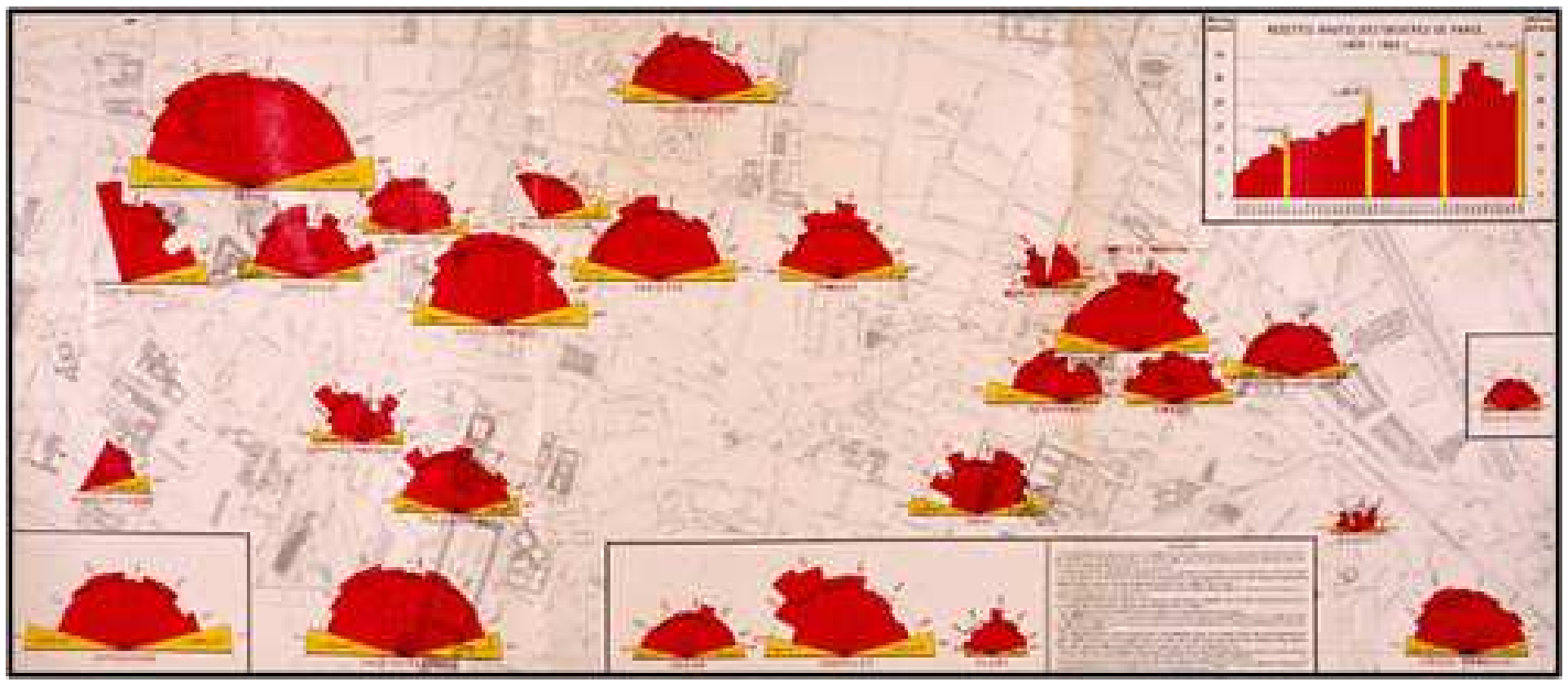

FIG. 14. *Polar-area diagrams. "Gross receipts of theaters in Paris from 1878 to 1889" (Exposition Universelle de 1889: Recettes brutes des théatres et spectacles de Paris 1878 à 1889). Each diagram uses sectors of length proportional to the receipts at a given theater in each year from 1878 to 1889, highlighting the values for the years of the Universal Expositions in yellow. Source: Album de Statistique Graphique, 1889, Plate 26.*

[18]For example, the time to travel from Paris to Toulouse was 330 hours in 1650; this had decreased to 104 hours by 1814 and to only 15.1 hours by 1887 with the development of railways. Today the same journey takes about 5 hours by the TGV route through Bordeaux. Montpellier and Marseilles, approximately the same geographical distance from Paris, became progressively closer in travel time. Today, the TGV gets you to Marseilles in about 3.25 hours.

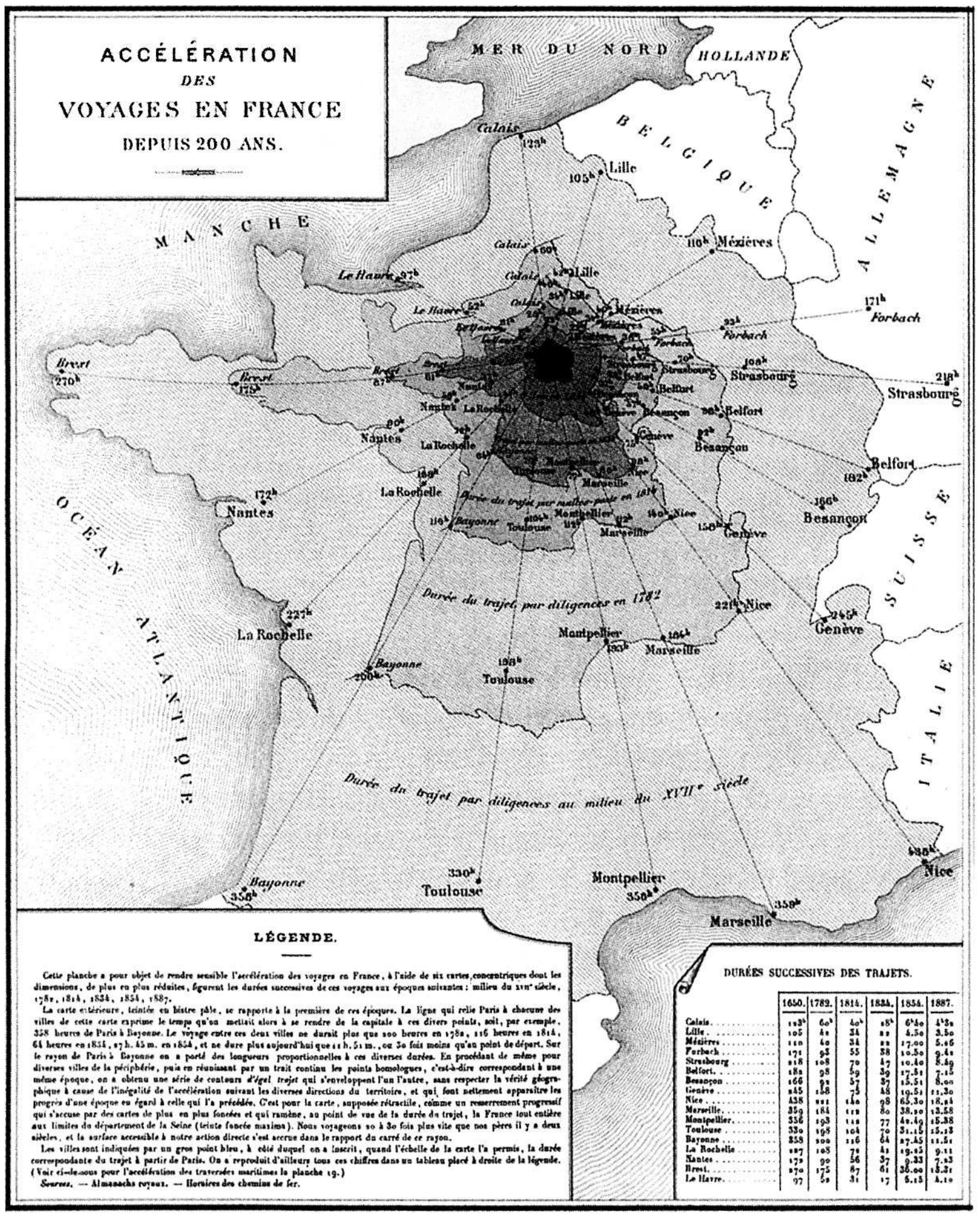

FIG. 15. *Anamorphic map: "Acceleration of Travel in France over 200 Years" (Accélération des voyages en France depuis 200 Ans). A set of five Paris-concentric maps scaled along the radial directions to show the relative decrease in travel time from 1789 to 1887. Source: Palsky (1996), Figure 63, Album de Statistique Graphique, 1888, Plate 8a.*

for example, Montpelier and Marseilles "moved" relatively closer to Paris than did Nice or Bayonne. The facing plate (Palsky, 1996, Figure 64) uses a similar anamorphic form to show the reduction in the price of travel from Paris to the same cities between 1798 and 1887.

Many of the plates used novel graphic forms to show changes over time of two or more related variables simultaneously on a schematic map of France. For example Figure 16 uses "planetary diagrams" to show two time-series of the transportation of principal merchandise by region over the years 1866–1894 in four-year intervals. The rays of the spiral are proportional to the average distance traveled; the diameters of the circles are proportional to tonnage moved.

Along with new graphic forms, there also had to be new ways to measure and count, as well as to show changes over time. Another pair of figures from the 1895–1896 volume (Plate 21, not shown) showed transportation on the national roads in France, calculated in "colliers réduits" (literally, "reduced collars"), a standard measure of transport adopted by the statistical bureau based on the equivalent number of animals for each conveyance. (This included tramways, but it is not clear how they counted bicycles and other modes of transport.) Both halves of this double plate use classed choropleth maps, but with a bipolar color scale, so that departments less than the mean (shaded red) contrast with those greater than the mean (shaded yellow). The left half of this plate shows the absolute numbers; the right half shows the difference in the values for 1894 compared with 1888.

The *Album de Statistique Graphique* was discontinued after 1897 due to the high cost of production.

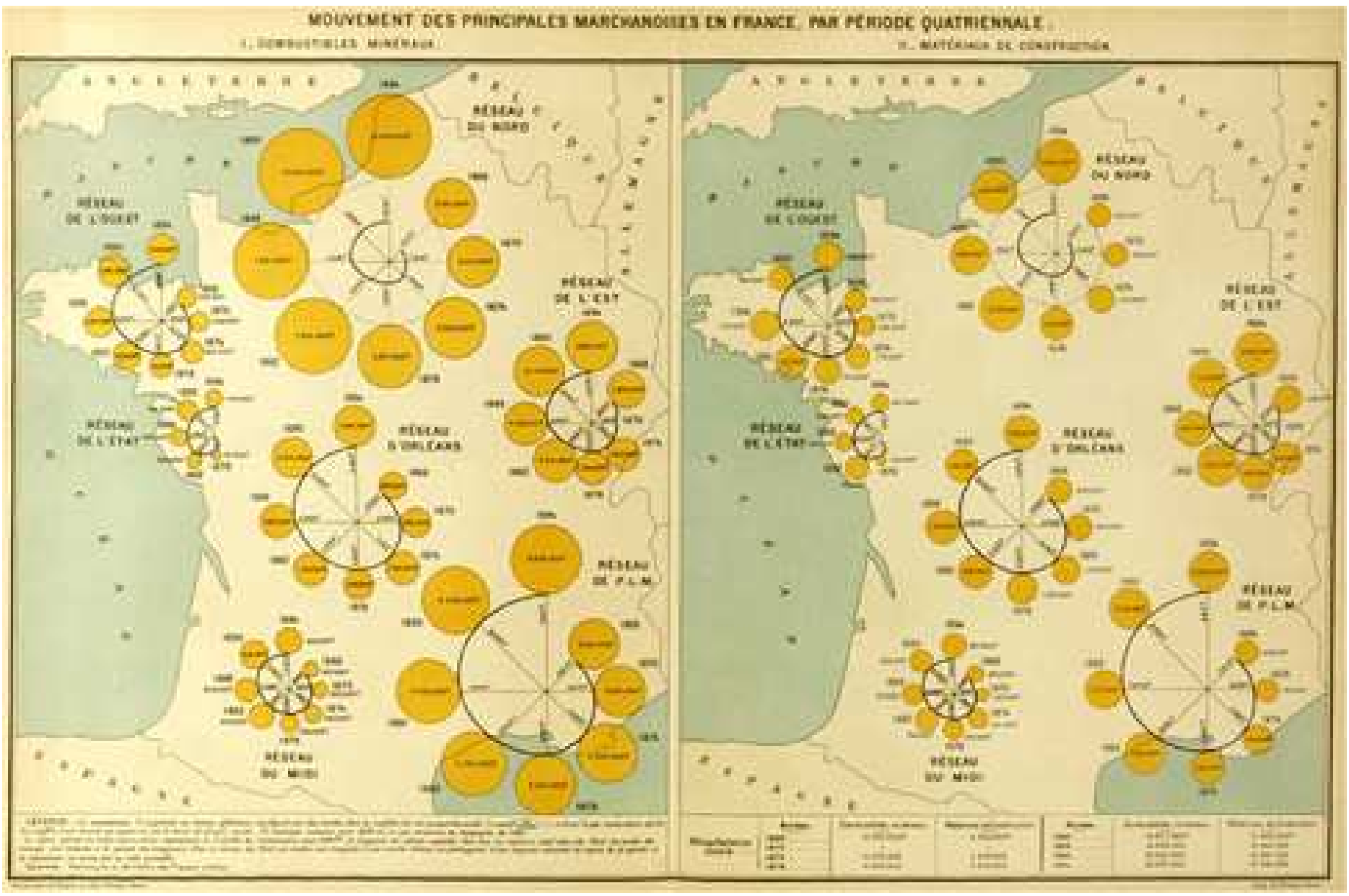

FIG. 16. *Planetary diagram: "Transportation of principal merchandise in France in Four Year Periods" (Mouvement des principales marchandises en France, par période quatriennale). Left: combustible minerals, for example coal, coke; right: construction materials. The length of rays indicate average distance; circle diameters represent tonnage moved. Source: Album de Statistique Graphique, 1895–1896, Plate 9.*

Faure (1918), page 295, called this decision a regrettable mistake and a serious loss to both administration and science.

3.3.2 *U.S. Census atlases.* Other striking examples representing high-points of the Golden Age may be found in the series of statistical atlases published by the U.S. Census Bureau in three decennial volumes for the census years from 1870 to 1890. The *Statistical Atlas of the Ninth Census*, published in 1874 under the direction of Francis A. Walker [1840–1897], was the first true U.S. national statistical atlas, composed as a graphic portrait of the nation. This was followed by larger volumes from each of the 1880 and 1890 censuses, prepared under the direction of Henry Gannett [1846–1914], sometimes described as the father of American government mapmaking.[19]

The impetus for this development stemmed largely from the expanded role given to the Census office following the U.S. Civil War. It was initially designed to serve the constitutional need to apportion congressional representation among the states. However, by June, 1872, the Congress recognized "the importance of graphically illustrating the three quarto volumes of the ninth census of the United States, by a series of maps exhibiting to the eye the varying intensity of settlement over the area of the country, the distribution among the several States and sections of the foreign population, and of the principal elements thereof, the location of the great manufacturing and mining industries, the range of cultivation of each of the staple productions of agriculture, the prevalence of particular forms of disease and other facts of material and social importance which have been obtained through such census" [Ex. Doc. No. 9, 42nd Congress] (Walker, 1874, page 1).

Accordingly, the atlas for the ninth census was composed of 54 numbered plates divided into three parts: (a) physical features of the United States: river systems, woodland distribution, weather, minerals; (b) population, social and industrial statistics: population density, ethnic and racial distribution, illiteracy, wealth, church affiliation, taxation, crop production, etc.; (c) vital statistics: age, sex and ethnicity distributions, death rates by age, sex, causes, distributions of the "afflicted classes" (blind, deaf, insane), etc. The plates were accompanied by eleven brief discussions of these topics, containing tables and other illustrations.

In carrying out his mandate, Walker stayed relatively close to his largely cartographic mission, but still found room to introduce novel graphic forms or redesign older ones to portray the American statistical landscape.

[19]Copies of these census atlases on CD, containing high-resolution, zoomable images, may be obtained from the Historic Print and Map Co., www.ushistoricalarchive.com.

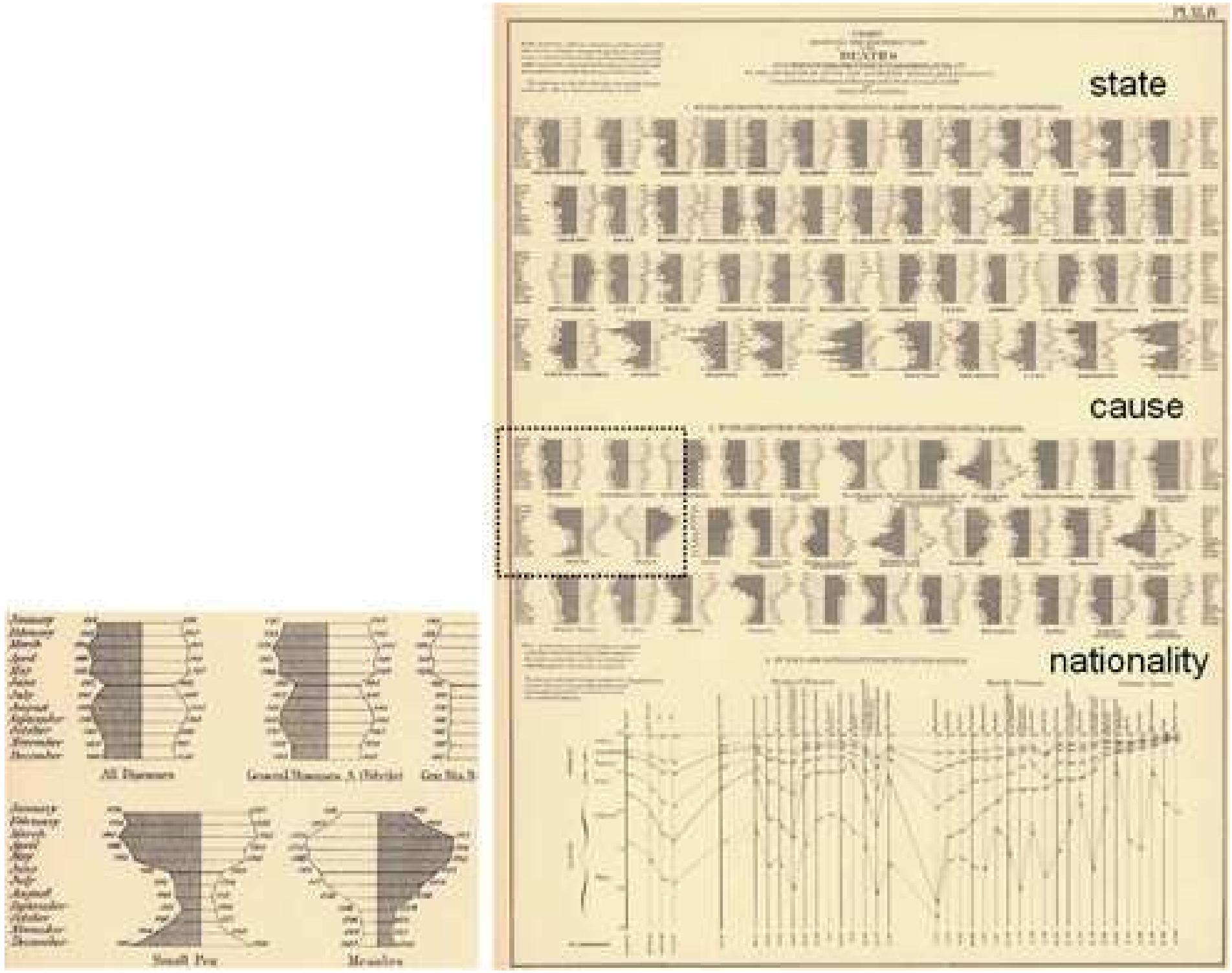

FIG. 17. *Bilateral histograms: "Chart showing the distribution of deaths ... by sex and month of death and according to race and nationality." Left: detail from causes of death; right: full plate, with labels for the three sections added. Source: Statistical Atlas of the United States, 1874, Plate 44.*

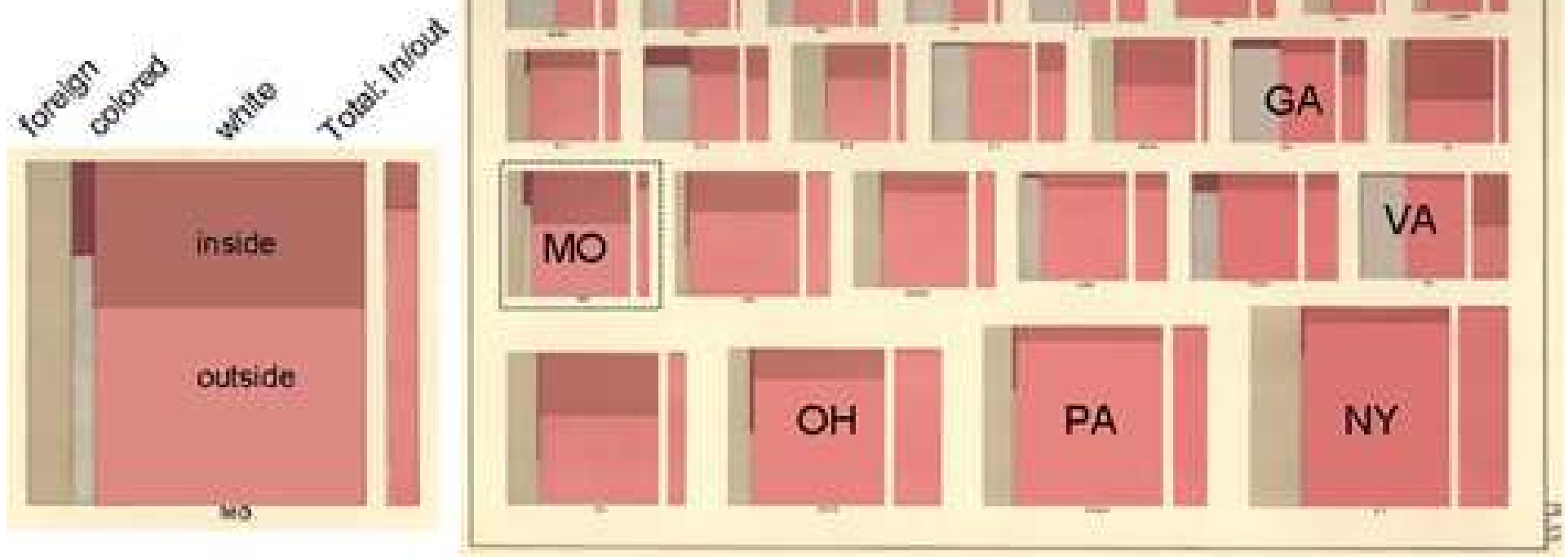

FIG. 18. *Mosaics/treemaps: Francis Walker, "Chart showing the principal constituents of each state" (1874). Left: detail for Missouri; right: full plate, with annotated labels added. Source: Statistical Atlas of the United States, 1874, Plate 20.*

Most noteworthy in this account is his development of the idea of showing two frequency distributions back-to-back, called generally a *bilateral histogram*, or an *age pyramid* when the classification is based on age. Figure 17 shows one particularly complex example that indicates the level of specificity the atlases attempted. Each bilateral histogram compares the number of deaths for males and females across months of the year; the sex which dominates is shaded. The top portion shows these for the U.S. states; these are arranged alphabetically, except for the last row, which contains small, largely western states. The middle portion shows these classified by cause of death. In the detail shown at the left, it can be seen most clearly that the shapes of these histograms vary considerably across diseases, and that some take their greatest toll on life in the winter months. The bottom portion is composed as a set of line graphs, classifying deaths according to nationality (i.e., native-white, colored, foreign-born) vertically and by age, groups of diseases, specific diseases and childhood diseases horizontally.

Another nice example (Figure 18) from the volume for the ninth census uses mosaic diagrams or treemaps to show the relative sizes of the state populations and the breakdown of residents as foreign-born, native-colored or native-white. The last two groups are subdivided according to whether they were born inside or outside that state, with a total bar for inside/outside added at the right. See Hofmann (2007) for a detailed reanalysis of this chart from a modern perspective. Other plates in this atlas (e.g., 31, 32) used similar graphic forms to show breakdowns of population by church affiliation, occupation, school attendance and so forth, but I view these as less successful in achieving their presentation goals.

Like Cheysson did in the *Albums*, Francis Walker (and his collaborators[20]) thought carefully about exactly what features of a topic should be emphasized in each display. Figure 19 shows one of four plates designed by Fred H. Wines, that closed the 1874 volume for the ninth census, with the goal of showing the distribution of the "afflicted classes" (blind, deaf mutes, insane, idiots) according to race, sex and nationality, while allowing for finer comparisons across other factors. The pie charts in the upper left and right corners give the total numbers of blind persons in the United States, divided by sex and nationality (native vs. foreign-born) on the left and divided by sex and race (white vs. colored) on the right. In the first horizontal row, the sex by nationality distribution (upper left) is shown separately for each state, with the area of each pie proportional to the total number in that state. The second horizontal row is a similar decomposition by state of the sex by race totals. The bottom row of circles attempts to show the increase in the number of blind from 1860 (inner circle) to 1870 (outer circle) for each state. With a consistent design for this series of plates, Walker was able to make the following summary description in his preface: "the males are shown to be in excess among the blind, the deaf mutes and the idiots; the females among the insane. The foreigners are shown to be in excess of their proportion among the blind and the insane; the natives among the deaf mutes and the idiots. ..." (Walker, 1874, page 3).

Following each of the subsequent censuses for 1880 and 1890, statistical atlases were produced with more numerous and varied graphic illustrations under the direction of Henry Gannett. These can be considered "the high-water mark of census atlases in their breadth of coverage, innovation, and excellence of graphic and cartographic expression" (Dahmann, 2001). The volume for the tenth census (Hewes and Gannett, 1883) contained nearly 400 thematic maps and statistical diagrams composed in 151 plates grouped in the categories of physical geography, political history, progress of the nation, population, mortality, education, religion, occupations, finance and commerce, agriculture, and so forth. The volume for the eleventh census (Gannett, 1898) was similarly impressive and contained 126 plates.

There were several summary illustrations using ranked parallel coordinate plots to compare states or cities over multiple dimensions or across all censuses. Figure 20, the final plate in the 1880 volume, compares the ranks of the states on 10 global measures, including population density, wealth, value of manufacturing, agriculture, live stock, taxation, education and illiteracy. The darkened bar in each column represents the average value for the nation as a whole. Among other things, the inverse relations between manufacturing and agriculture, and between taxation and education are striking. Figure 21 was one of the initial plates in the 1890 atlas, showing the relative rankings of state populations from all censuses, 1790–1890, using color-pattern coding of the state symbols to aid readability.

[20]The title page for this *Statistical Atlas* says, "with contributions from many eminent men of science, and several departments of the government." The volume contained 11 memoirs and discussions on these topics, signed individually.

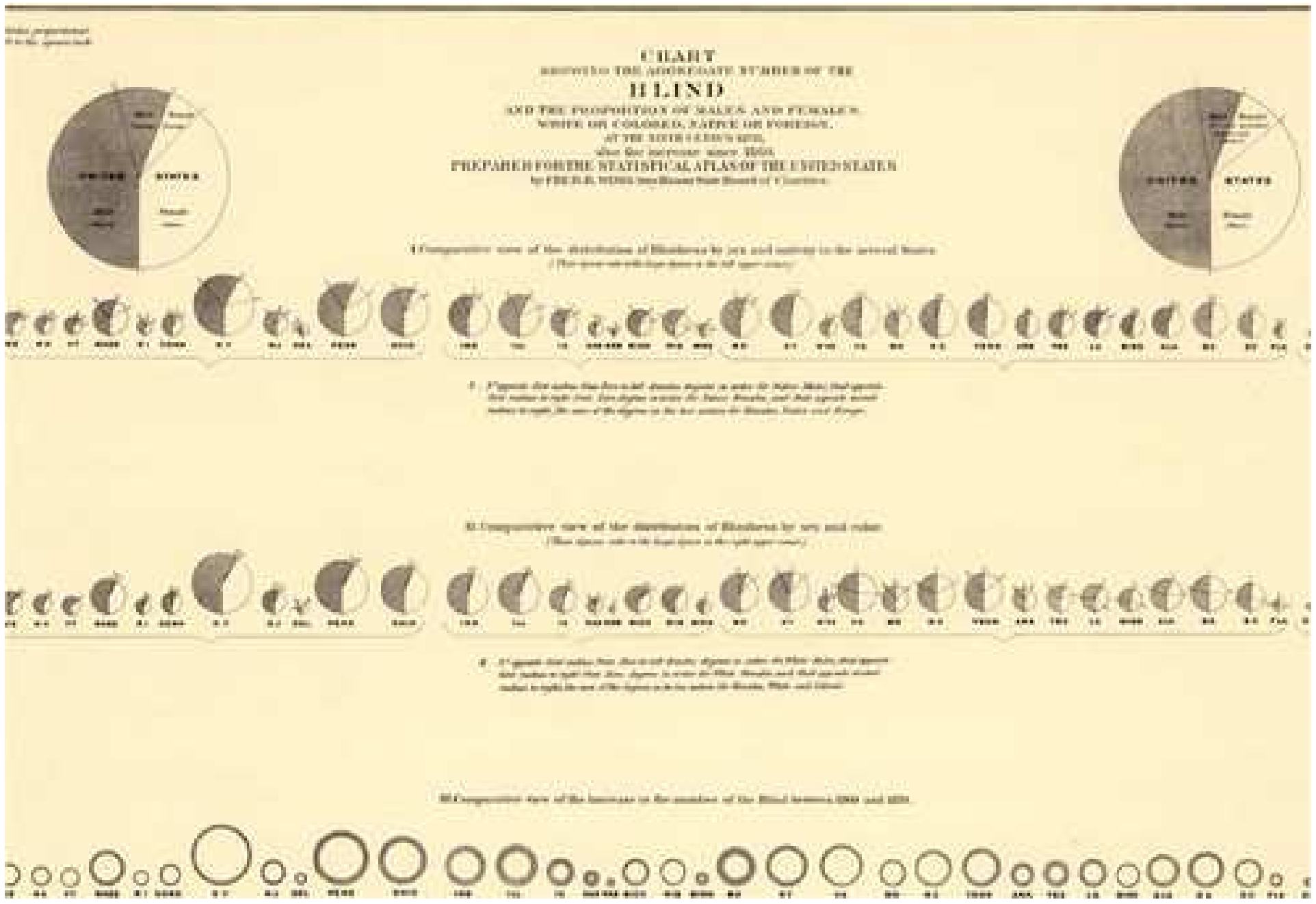

FIG. 19. *Proportional pie and circle diagrams: Fred H. Wines, "Chart showing the aggregate number of the blind and the proportion of males and females, white or colored...," (1874). Source: Statistical Atlas of the United States, 1874, Plate 51.*

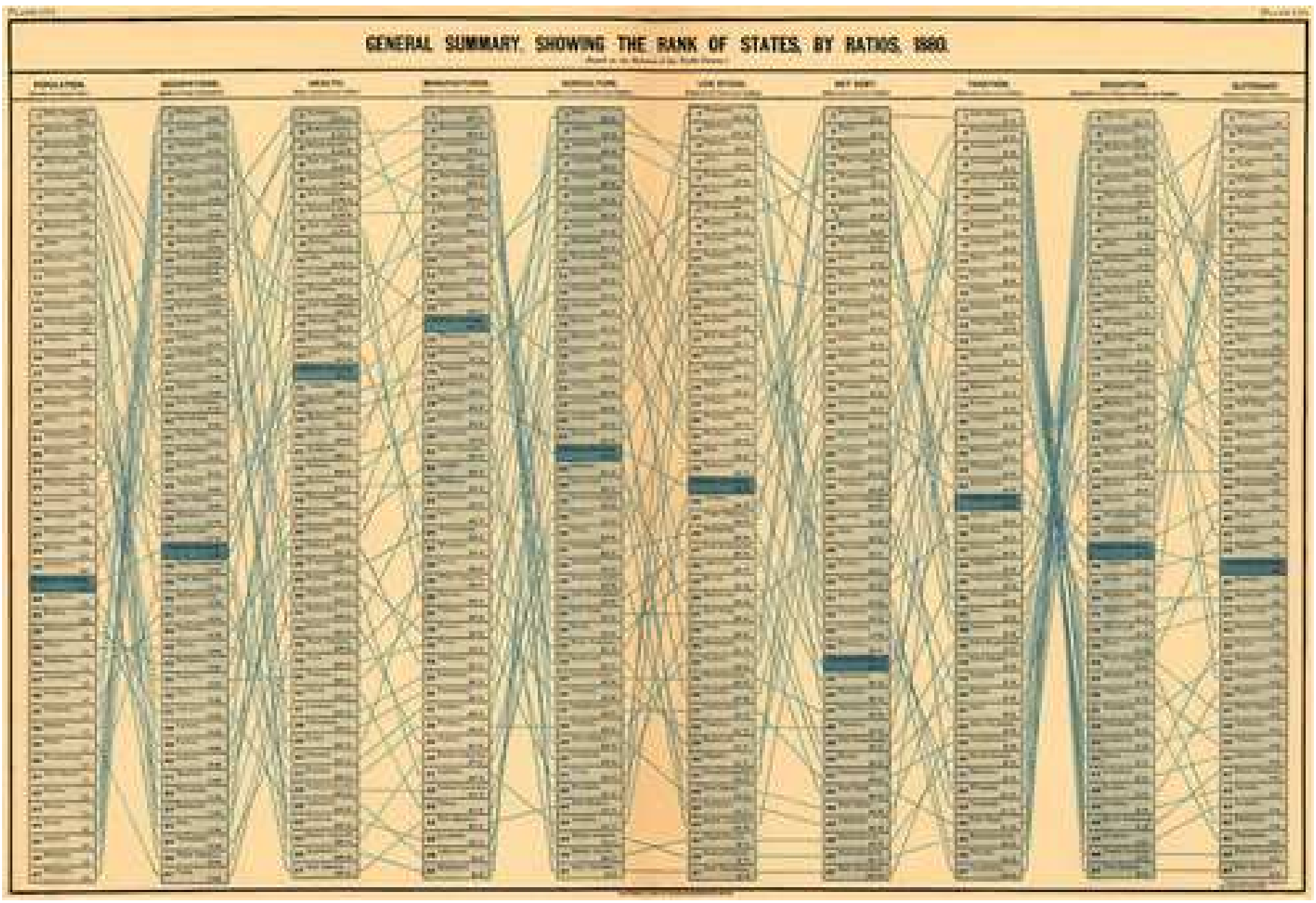

FIG. 20. *Multivariate ranked-list, parallel-coordinates chart: Henry Gannett, "General Summary, Showing the Rank of States, by Ratios, 1880." Source: Statistical Atlas of the United States, Hewes and Gannett (1883), Plate 151.*

In addition, these volumes are notable for the use of trellis-like collections of shaded maps showing the rate or density of interstate migration, religious membership, deaths from known causes, and so forth, but in a way that allowed easy comparison across the stratifying variables. Other plates used well-designed combinations of different graphic forms (maps, tables, bar charts, etc.) to show different aspects of a topic. See Figure 22.

3.3.3 *Other statistical atlases.* There were other statistical atlases and albums which deserve mention in this account and which reflect the aspirations, enthusiasm for graphics and excellence of design that characterize the Golden Age. For reasons of space, my coverage of these will be relatively brief, which makes it even harder to choose a few examples for this discussion.

Two statistical albums were produced in Switzerland, in conjunction with public expositions in Geneva and Berne in 1896 and 1914, respectively (Statistischen Bureau, 1897, 1914).[21] Together, these contained 92 plates, classified by topics including politics, area, population (e.g., births, deaths by various causes, marriage), transportation, industry and commerce, and so forth. Similarly to the U.S. Census atlases, the Swiss albums used a wide variety of graphical forms (line graphs, bar charts, pie diagrams, proportional squares, shaded maps using unipolar and bipolar color scales, etc.). Many of these combined several different graphical methods to provide views of different aspects or breakdowns of a given topic.

For example, Figure 23, from the 1897 album, uses a central pie chart surrounded by bar charts to show causes of death in Switzerland from 1890 to 1894. It is an early example of the idea of linking separate displays by a perceptual attribute—here, the color of the pie slices, or, what one might do today using drill-down to show finer detail. The large black slice at the bottom is labeled "Other causes," but is mostly death by suicide; the bar graphs at the bottom left and right show the distributions of suicide by month of the year and by method, respectively. From the bar graphs one can see that most people committed suicide by hanging, followed by firearms, and that (somewhat inexplicably), suicides were most common in the summer months.

[21] These are available on CD, in a form that integrates the explanatory text with the plates, from the Swiss Bundesamt für Statistik, http://www.bfs.admin.ch/bfs/portal/fr/index/news/publikationen.html?publicat, order number 760-0600-01.

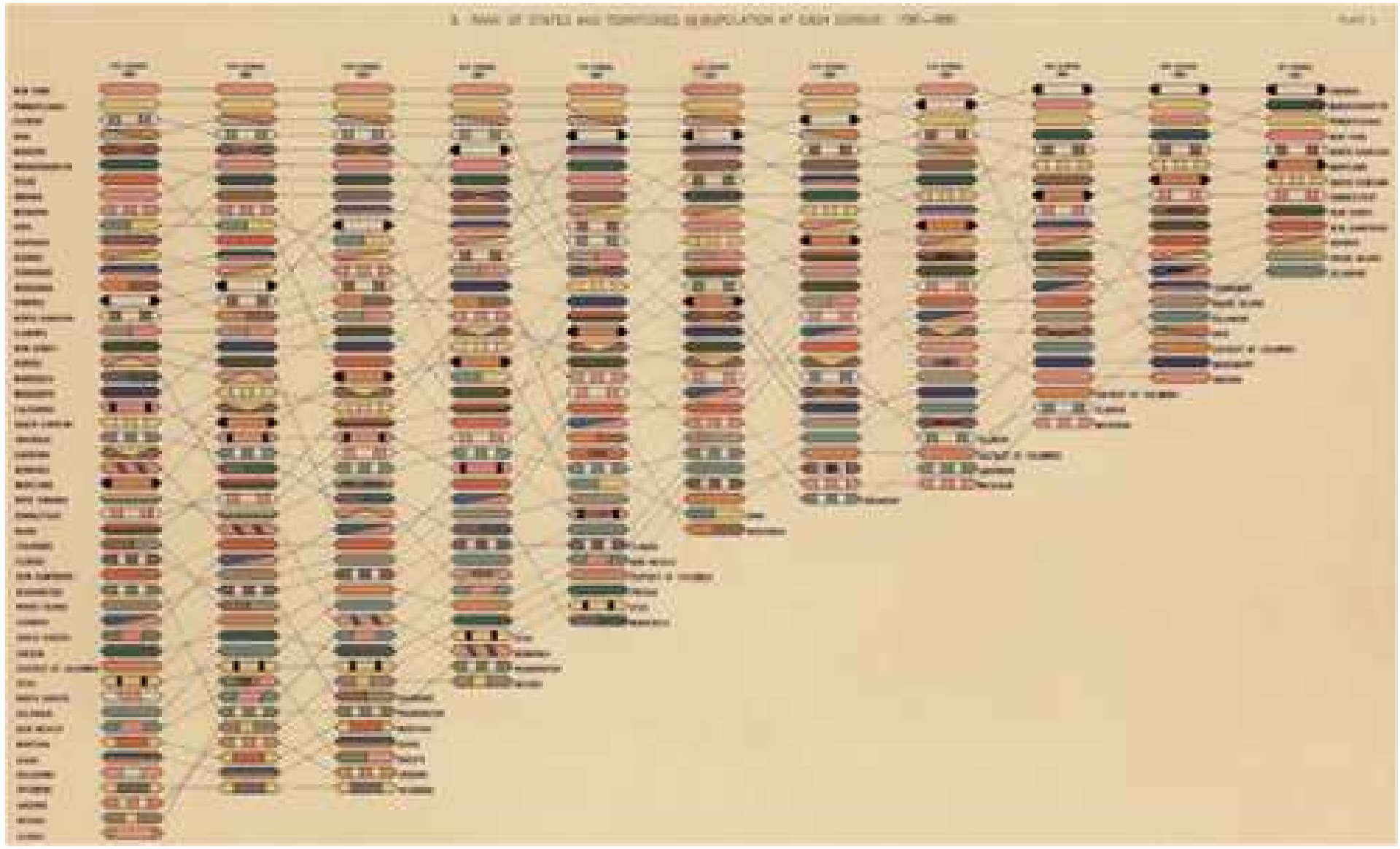

Fig. 21. *Linked parallel-coordinates time-series diagram: Henry Gannett, "Rank of States and Territories in Population at Each Census, 1790–1890" (1898). Source: Private collection, Statistical Atlas of the United States, Gannett (1898), Plate 2.*

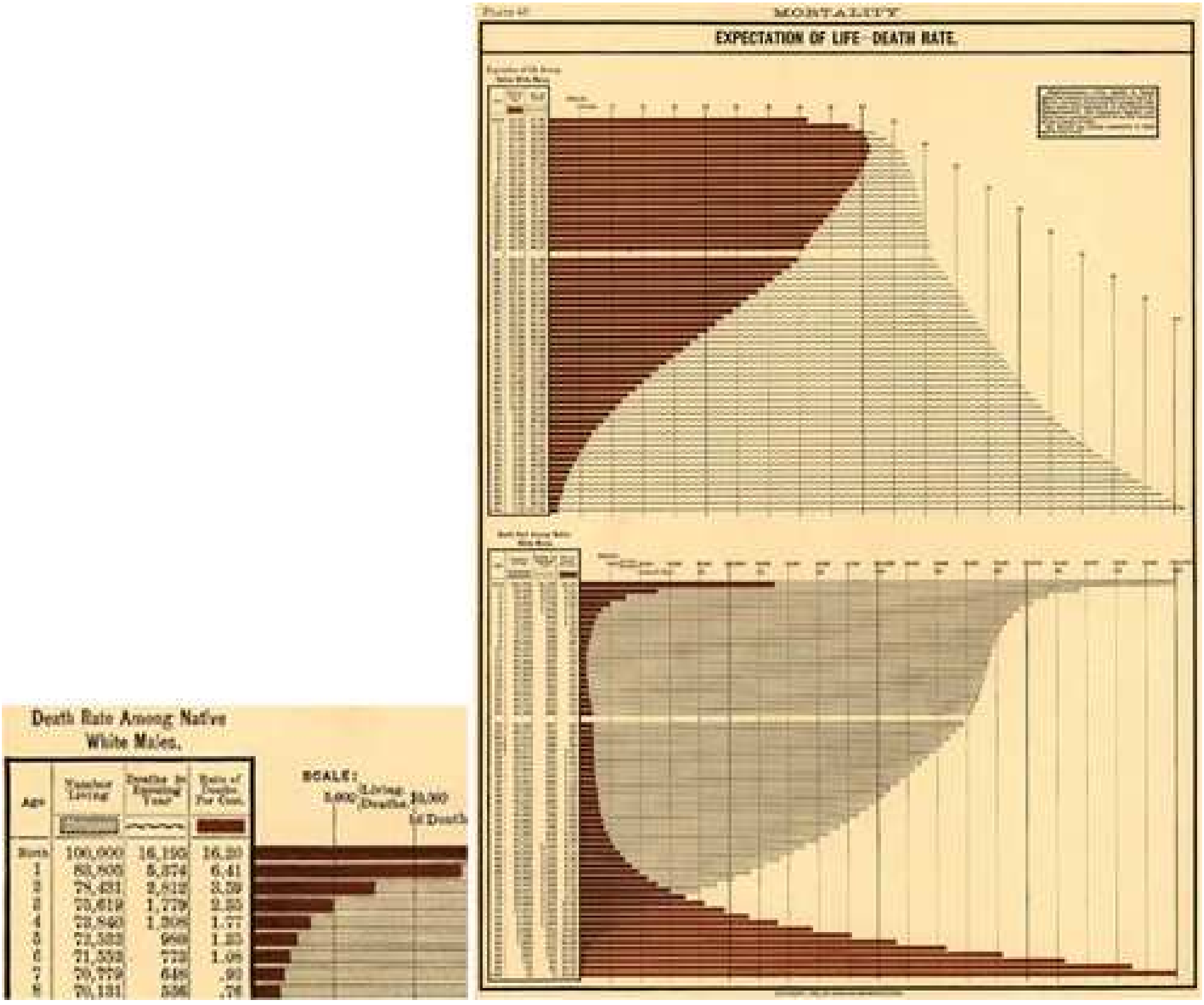

FIG. 22. *Multifunction bar and line graphs: Mortality: life expectancy and death rates by age, among native white males. Left: detail from death rates; right: full plate. Source: Statistical Atlas of the United States, 1880, Plate 40.*

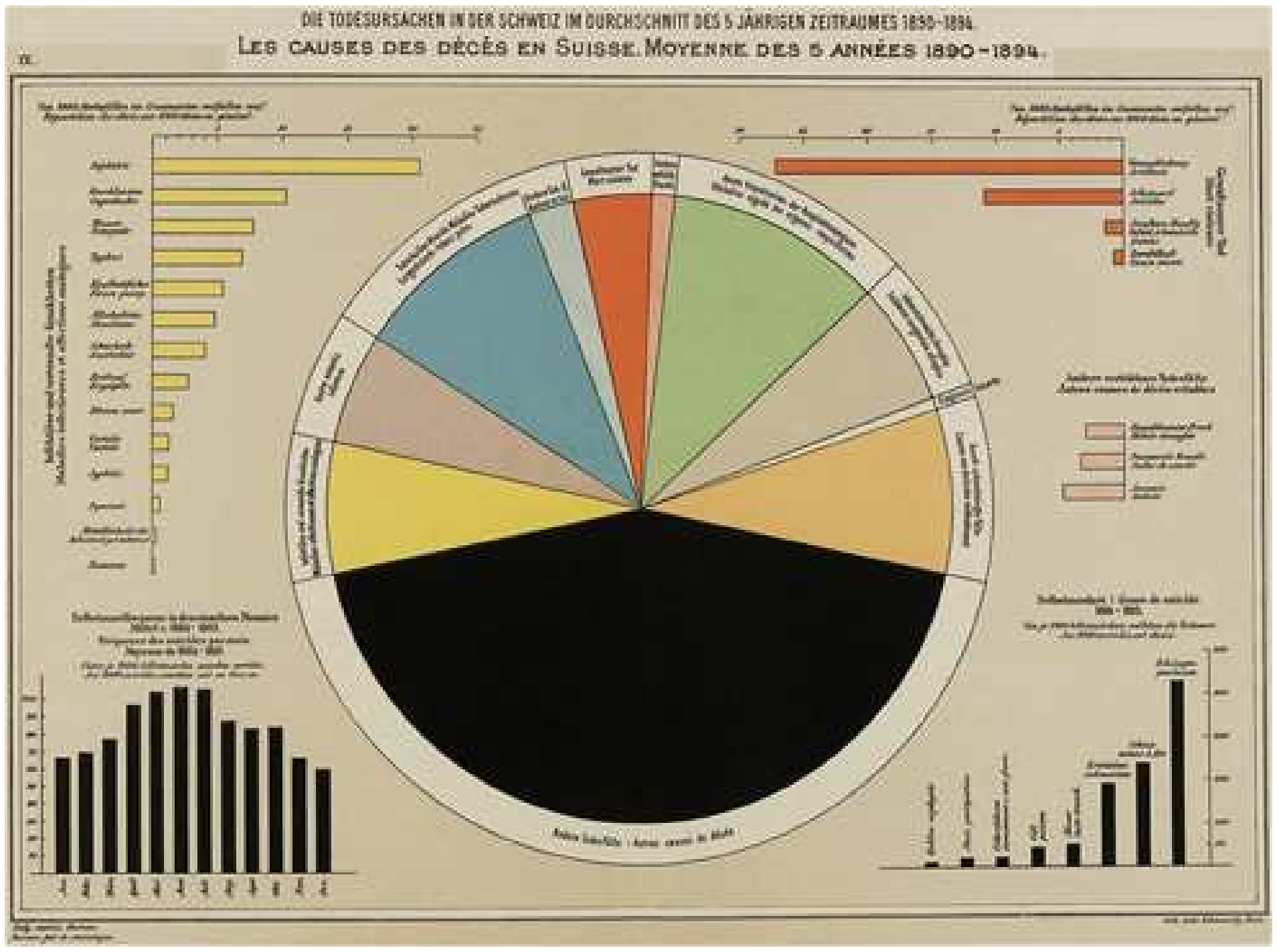

FIG. 23. *Color-linked pie and bar charts: Swiss Statistical Office, "Causes of Death in Switzerland, Average of 5 Years, 1890–1894" (Les causes des décés en Suisse Moyenne des 5 années 1890–1894). Some of the causes shown in the pie chart are color-linked to decompositions shown in the bar charts. Source: Private collection, Switzerland Graphical Statistical Atlas, 1897, Plate 9.*

The yellow sector, representing infectious diseases, is color-linked to the bar graph at the top left, showing diphtheria as the most common fatal disease.

Figure 24 is another example, of a type we have not seen before. The presentation goal is to show the improvement in the scores of military recruits from 1880 to 1912 on an examination based on reading, composition, mathematical calculation and civil knowledge. Each part was scored from 1–5 (1 = best), so the total ranged from 4 to 20. The top portion of the graphic shows the distribution of the Swiss cantons' scores in 1880 and 1912 along a 4–20 scale. It is readily apparent that the average score improved considerably and also that the variability decreased over these 32 years. The bottom portion gives a more detailed comparison, using two sets of back-to-back histograms (left: data from 1886; right: 1912). On each side, the top histogram shows the proportion of recruits in each canton achieving "good" results (marks of 1 in at least two areas); the bottom shows the proportion earning "bad" results (≥4 in two or more areas). As in Figure 23, the intent was to tell a graphic story with a clear main message and to provide additional detail.

In this period, national statistical atlases were also occasionally produced elsewhere in Europe (Russia: 1873; Germany: 1876–1878; Austria: 1882–1885), and local atlases were created in Paris, Frankfurt, Berlin and elsewhere. Among these, the two volumes by Jacques Bertillon in the *Atlas de statistique graphique de la ville de Paris* (Bertillon, 1889, 1891) deserve mention as one facet of an ambitious attempt to portray the city of Paris graphically in terms that would allow its functions and structures to be studied scientifically and contribute to more efficient management (Picon, 2003). Once again, the great variety of topics selected for study, and the graceful blending of a wide range of graphical methods employed, cannot fail to impress. Bertillon's statistical atlases of Paris are described in detail in Palsky (2002b) and nine of his plates appear in Picon and Robert (1999).

Figure 25 shows a novel graphic form invented by Bertillon to show two variables and their product simultaneously by the width, height and area of rectangles. The subject concerns the loans made by the offices of the Mont de Piété, a system of public pawn shops authorized by the government for making low-interest loans available to the poor (now the Crédit Municipal de Paris). It was first established in 1777 and by 1880, had 26 offices throughout the city. For each office, the widths of the rectangles show the number of loans, and the heights show the average amount; thus, the area represents the total value of loans. The three rectangles show, respectively, new loans (left, shaded pink), renewals (right, yellow) and redemptions (middle, cross-hatched). The main office of the Mont de Piété is the largest rectangle (in the Marais, near Place des Vosges). One can see that most bureaus made a great many loans of small value; however, one office (near Gare du Nord) reversed this pattern with a small number of loans of large value.

Bertillon's statistical atlases had been planned as a continuing series, but, like the *Album de Statistique Graphique*, were discontinued for financial reasons. The age of state-sponsored statistical bureaus with a mandate to produce such graphic volumes had begun to wane. After the first World War, a few more graphical statistical atlases were published in emerging countries (e.g., Latvia, Estonia, Romania, Bulgaria) as a concrete symbol of national affirmation and a step in the construction of national identity. The Golden Age, however, had come to a close.

## 4. END OF THE GOLDEN AGE: THE MODERN DARK AGES

If the last half of the 19th century can be called the Golden Age of Statistical Graphics, the first half of the 20th century can equally be called the "modern dark ages" of data visualization (Friendly and Denis, 2000). What had happened?

### 4.1 Demise

The enthusiasm for graphical analysis that characterized government-sponsored statistical atlases of the Golden Age was curtailed for a number of reasons. There is a longer history of this period, and the general decline shown in Figure 1 was not without some notable success stories. For the present account, two main factors stand out in relation to graphical excellence of statistical albums and graphical innovation.

First, those with local or time-limited purposes (such as the Swiss albums, Bertillon's statistical atlases of Paris, and others not mentioned here) were simply allowed to expire. More general series, including the *Album de Statistique Graphique* and others, eventually succumbed to the high cost of production (Faure, 1918; Palsky, 1996). As witness to this, the final two U.S. Census statistical atlases, published after the 1910 and 1920 censuses, "were both routinized productions, largely devoid of color and graphic imagination" (Dahmann, 2001).

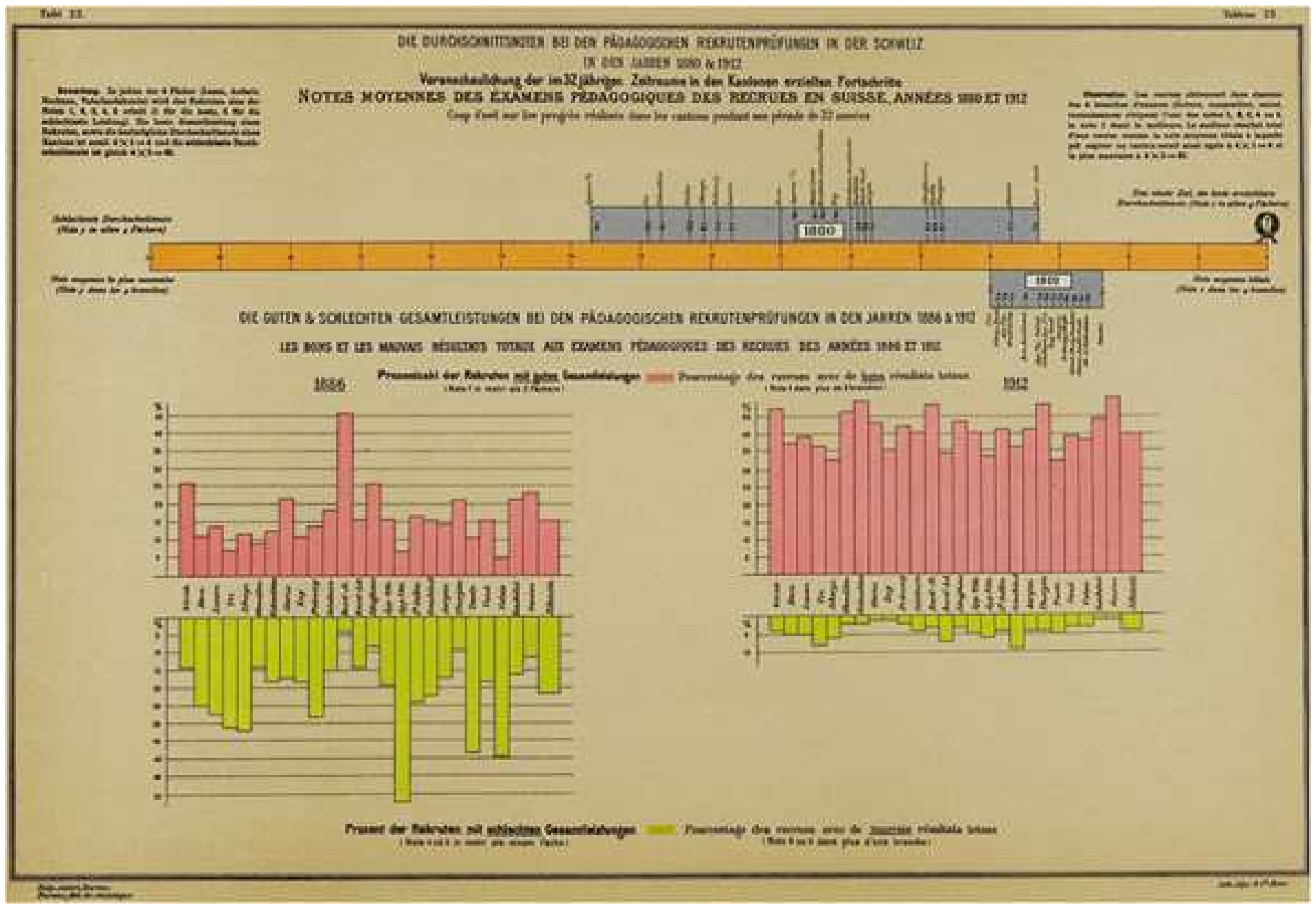

Fig. 24. *Proportional area diagram: Swiss Statistical Office, "Mean scores on examinations of recruits in Switzerland, 1880 and 1912" (Notes moyennes des examens des recrues en Suisse, Années 1880 et 1912). Source: Private collection, Switzerland Graphical Statistical Atlas, 1914, Plate 23.*

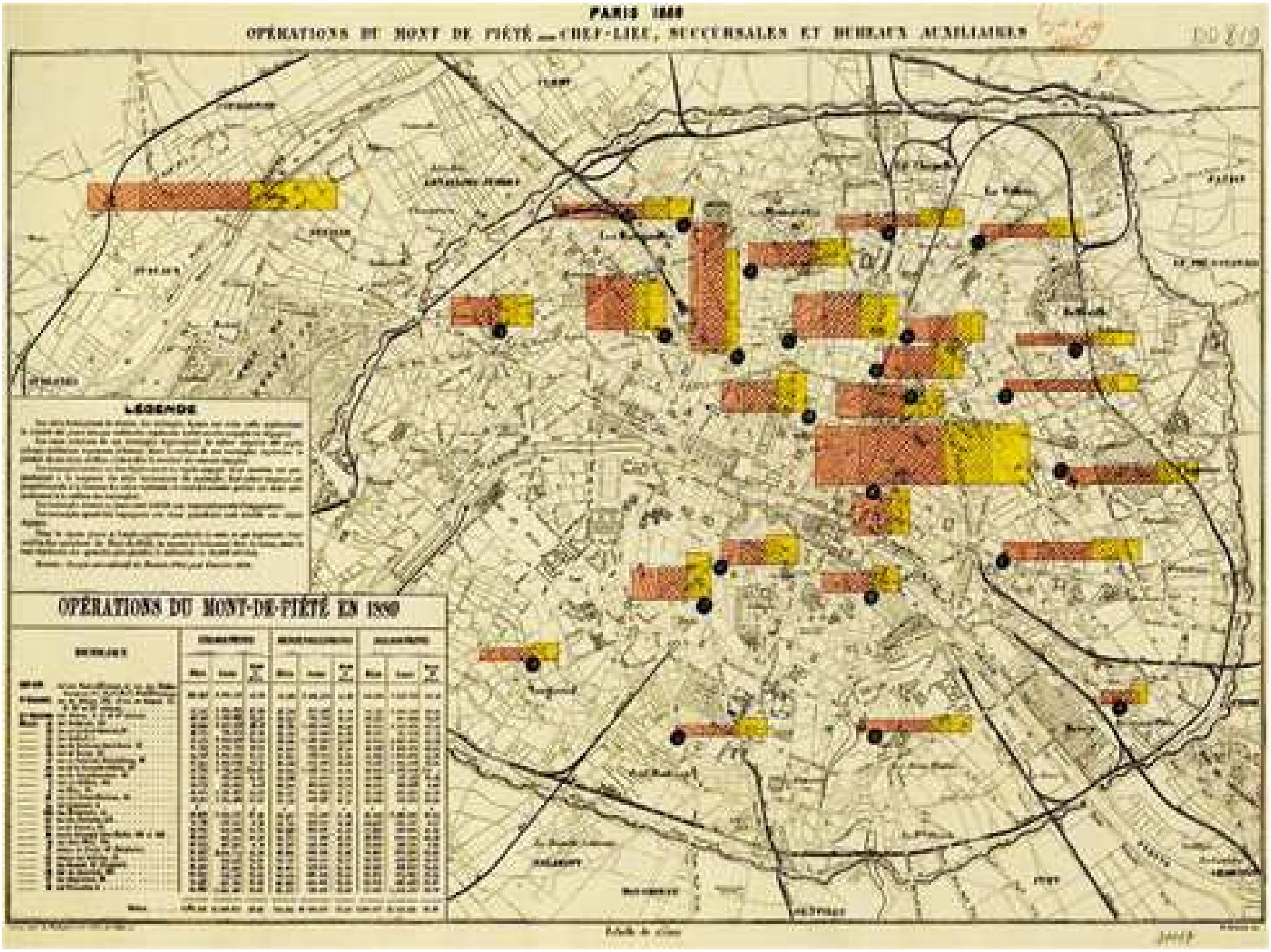

Fig. 25. *Three-variable rectangular cartogram: Jacques Bertillon, "Operations of the Monts de piété." For each office of the Mont de Piété, the width of the rectangles shows the number of loans, and the height shows the average amount, so the area represents the total value of loans. Bertillon (1891), Plate 2. Source: Picon and Robert (1999).*

There were a few notable exceptions to this decline in official uses of graphical methods after 1910. In the negotiations leading to the Paris Peace Conference after World War I, thematic maps—particularly those showing ethnic composition—played an important and sometimes decisive role in redrawing national boundaries in Central Europe and the Balkans. Palsky (2002a) provides a full discussion of this impact, with a focus on the contributions to redefining the territory of Romania by Emmanuel de Martonne [1873–1955] (secretary of the French committee of geographical experts set up in 1915, and one of the leading physical geographers of the early 20th century).[22]

A second, potentially more powerful, force had also been brewing. This would turn the attention and enthusiasm of both theoretical and applied statisticians away from graphic displays back to numbers and tables, with a rise of quantification that would supplant visualization. This point has been argued before (Friendly and Denis, 2000) and will only be slightly extended here.

The statistical theory that had started with games of chance and the calculus of astronomical observations developed into the first ideas of statistical models, starting with correlation and regression, due to Galton, Pearson and others. By 1908, W. S. Gosset developed the $t$-test, and between 1918–1925, R. A. Fisher elaborated the ideas of ANOVA, experimental design, likelihood, sampling distributions and so forth. Numbers, parameter estimates—particularly those with standard errors—came to be viewed as precise. Pictures of data became considered—well, just pictures: pretty or evocative perhaps, but incapable of stating a "fact" to three or more decimals. At least it began to seem this way to many statisticians and practitioners.[23]

In the social sciences, Gigerenzer and Murray 1987 referred to this as the "inference revolution"—the widespread adoption of significance testing as the essential paradigm for doing empirical research, and getting it published. In economics, time-series graphs had been dominant in the 19th century (Klein, 1997; Maas and Morgan, 2005) but were displaced by more modern analytic techniques; as a result, "graphs could be considered at best unnecessary, and at worst seriously misleading" (Morgan, 1997, page 76). See, for example, Cleveland (1984) and Bestetal (2001) for empirical studies on the use of graphs in various disciplines.

As a result of these changes in resources and interest, there were hardly any new graphical innovations in this period,[24] and few examples of graphical excellence. Yet, there were a few developments that bear mention in relation to the end of the Golden Age. First, statistical graphics became popularized and entered the main stream. A spate of English textbooks began to appear, for example, Bowley (1901), Peddle (1910), Haskell (1919), Karsten (1925), college courses on graphics were developed (Costelloe,

[22]To appreciate the controversial and often political aspects of this debate, it is worth quoting the cynical sentiments of Isaiah Bowman, chief geographical expert from the American delegation, on the role played by statistical maps:

> Each one of the Central European nationalities had its own bagful of statistical and cartographical tricks. When statistics failed, use was made of maps in color. ... A map was as good as a brilliant poster, and just being a map made it respectable, authentic. ... It was in the Balkans that the use of this process reached its most brilliant climax. (Bowman, 1921, page 142.)

Bowman's views, based more on political desires to reshape Europe to the advantage of U.S. interests, were publicly condemned (e.g., Anonymous, 1921). Nevertheless, the impact of ethnographic statistical maps for both propaganda and rational debate had been highlighted in this episode.

[23]Of course, Galton, Pearson and even Fisher were enthusiastic graph people, but many of those who extended their ideas (e.g., F. Y. Edgeworth, G. U. Yule and even Pearson himself) turned their attention to mathematical and analytic aspects of correlation, regression and general theories of statistical distributions. Later developments of likelihood-based inference, decision theory and measure theory served to increase the sway of more formal mathematical statistics. "Indeed, for many years there was a contagious *snobbery* against so unpopular, vulgar and elementary a topic as graphics among academic statisticians and their students" (Kruskal, 1978, page 144, italics in original).

[24]In this time, one of the few graphical innovations—one perhaps best forgotten—was the development of pictograms by Willard C. Brinton (1914) and later, Otto Neurath in 1924 (see Neurath, 1973) in which numerical values were shown using schematic icons stacked in bar graph form, or with "size" proportional to value. The goal was to make numerical results more comprehensible by the general population. But graphic designers often got the details wrong, creating a large supply of exemplars of "bad graphs" (Tufte, 1983b; Wainer, 1984). Brinton himself (Brinton, 1914, Figure 39) explicitly cautioned against some of these abuses, but they live on daily in the pages of *USA Today* and similar publications.

1915; Warne, 1916), and statistical charts, perhaps mundane, entered standard use in government (Ayres, 1919) and commerce.

Second, in this period, graphical methods proved crucial in a number of new insights, discoveries, and theories in astronomy, physics, biology and other natural sciences. Among these, one may refer to (a) E. W. Maunder's (1904) "butterfly diagram" to study the variation of sunspots over time, leading to the discovery that they were markedly reduced in frequency from 1645 to 1715 (the "Maunder minimum"); (b) the Hertzsprung–Russell diagram (Hertzsprung, 1911; Spence and Garrison, 1993), a log–log plot of luminosity as a function of temperature for stars, used to explain the changes as a star evolves and laying the groundwork for modern stellar physics; (c) the discovery of the concept of atomic number by Henry Moseley (1913) based largely on graphical analysis: a plot of square-root of the frequencies of certain spectral lines vs. rank-order in Mendeleev's periodic table; and (d) the discovery in macroeconomics of the inverse relation between inflation and unemployment over time, based on what would be called the "Phillips curve" (Phillips, 1958). See Friendly and Denis (2005) for more detailed discussion of these developments and a few insights resulting from a modern reanalysis of some of these early scatterplots. In general, the *use* of graphical methods in the natural sciences continued throughout this period, though no new ground was broken.

### 4.2 Rebirth and Impact

In Figure 1, the periods from about 1950–1975 (labeled "Rebirth") and from 1975 to the present ("High-D Vis") are shown as an upswing in the density of contributions to statistical graphics that are cataloged as significant events in the Milestones Project. It is out of place here to fully recount the developments that led to the rebirth of interest and the explosive growth in truly novel methods of high-dimensional data visualization and dynamic, interactive graphics in recent times. Some of this is covered in Friendly (2005) and in Friendly (2008), Sections 2.7–2.8,[25] but the most modern history of this area is still unfolding.

In this article, it *is* useful to ask whether the innovations of graphic design and technique or the scope and excellence of execution by which I characterize the Golden Age had any real impact on the rebirth of interest in graphical methods or on the practices and techniques of today. I believe it is fair to say that this past work had little direct influence, at least initially, largely because it was unknown to researchers and graphic developers.

New attitudes toward graphics from the 1950s through the 1970s soon gave rise to a spate of apparently novel graphical methods: metroglyphs Anderson (1957), star plots (Siegel, Goldwyn and Friedman, 1971), fourfold circular displays for $2 \times 2$ tables of counts (Fienberg, 1975b). Slightly later came the ideas behind mosaic displays (Hartigan and Kleiner, 1981) and parallel coordinates plots (Inselberg, 1985). However, these appeared to have been developed *de novo*, without any reference to similar ideas from past history. All the "new" ideas I just mentioned had in fact been invented in the 19th century, particularly in various statistical atlases reviewed in Section 3.3.

The new interest prompted some to read the histories of the graphic method by Funkhouser (1937) and even Marey (1885); shortly, there appeared new historical bibliographies (Feinberg and Franklin, 1975), reviews (Beniger and Robyn, 1978; Fienberg, 1975a; Wainer and Thissen, 1981) or general presentations (Tufte, 1983b) on statistical graphics, that attempted to connect present methods with those of the past.

In summary, to answer the question of "what impact?," I would say that many, if not most, of the widely used static graphical methods and ideas of today have their origin in the later part of the 19th century, but these roots have only been explored in detail more recently. I cite a few examples here (some mentioned earlier) with references to historical studies where available:

- scatterplots and scatterplot smoothing, going back to Herschel (1833); see Friendly and Denis (2005), Hankins (2006)
- parallel coordinates plots (Lallemand, 1885; d'Ocagne, 1885)
- mosaic displays for cross-classified tables, originating with Mayr (1874); see Ostermann (1999) and Friendly (2002a)

[25]Very briefly, the revival of interest in statistical graphics may be said to stem from, among other factors, (a) John W. Tukey's (Tukey, 1962) prophetic call for attention to data analysis and graphics, followed by a host of simple, yet novel, methods that have now become mainsteam (Tukey, 1972, 1977) and (b) Jacques Bertin's *Semiologie Graphique* (Bertin, 1967), that established a rational structure for organizing the visual and perceptual elements of graphics according to features and relations in data.

- polar-area charts and star charts, originating with Guerry (1829b) and Nightingale (1857)
- clustered heatmaps, widely claimed as the invention of molecular biologists (Weinstein, 2008), but which go back to a shaded table by Toussaint Loua (1873) and Jacques Bertin's (1967) "reorderable matrix." See Wilkinson and Friendly (2008) for this history.

## 5. CONCLUSIONS

The principal objective of this paper has been to tell the story of a particular slice of the history of statistics and data visualization, the Golden Age, that deserves to be recognized—even revered—for the contributions it made to statistical thought and practice in that time and also for the legacy it provides today. I have tried to show how a collection of developments in data collection, statistical theory, visual thinking, graphic representation, symbolism from cartography and technology combined to produce a "perfect storm" for statistical graphics and thematic cartography in the last half of the 19th century. Moreover, many of these early statistical maps and diagrams, drawn by hand in the pre-computer era, were able to achieve a high degree of graphical impact—what Tukey (1990) referred to as interocularity (the message hits you between the eyes). This is a lesson to bear in mind as we go forward.

## ACKNOWLEDGMENTS

This work is supported by Grant 8150 from the National Sciences and Engineering Research Council of Canada. I am grateful to the archivists of many libraries and to *les Chevaliers des Album de Statistique Graphique*, in particular: Antoine de Falguerolles, Ruddy Ostermann, Gilles Palsky, Ian Spence, Stephen Stigler, Antony Unwin and Howard Wainer for historical information, images and helpful suggestions. Editorial comments by Yvonne Lai, Lee Wilkinson, Waldo Tobler, several reviewers and the editors helped considerably in improving the manuscript.

### Notes for the references

A few items in this reference list are identified by shelfmarks or call numbers in the following libraries: BL: British Library, London; BNF: Bibliothèque Nationale de France, Paris (Tolbiac); ENPC: École Nationale des Ponts et Chaussées, Paris; LC: Library of Congress; SBB: Staatsbibliothek zu Berlin.